\documentclass[12pt,preprint]{aastex}

\begin{document}

\title{Gravitational Waves from Phase-Transition Induced Collapse 
of Neutron Stars}

\author{L.-M. Lin\altaffilmark{1,2,3,4},
K. S. Cheng\altaffilmark{2}, 
M.-C. Chu\altaffilmark{3}, 
and W.-M. Suen\altaffilmark{2,3,4} }

\altaffiltext{1}{Laboratoire de l'Univers et de ses Th\'{e}ories,
Observatoire de Paris, F-92195 Meudon Cedex, France.}

\altaffiltext{2}{Department of Physics, University of Hong Kong,
Hong Kong, China.}

\altaffiltext{3}{Department of Physics, Chinese University of 
Hong Kong, Hong Kong, China.}

\altaffiltext{4}{McDonnell Center for the Space Sciences,
Department of Physics, Washington University, St. Louis, MO 63130.}

\date{13 March 2006}

\begin{abstract}  
We study the gravitational radiation from gravitational collapses of
rapidly rotating neutron stars induced by a phase transition from normal
nuclear matter to a mixed phase of quark and nuclear matter in the core of the
stars. The study is based on self-consistent three-dimensional
hydrodynamic simulations with Newtonian gravity and a high-resolution 
shock-capturing scheme and the quadrupole formula of gravitational radiation.  
The quark matter of the mixed phase is described by the MIT bag model and 
the normal nuclear matter is described by an ideal fluid EOS.
While there is a broad range of interesting astrophysics problems associated with 
the phase-transition-induced gravitational collapse, we focus on the following:  
First, we determine the magnitudes of the emitted gravitational 
waves for several collapse scenarios, with gravitational wave amplitudes 
ranging from $0.3 \times 10^{-22}$ to $1.5\times 10^{-22}$ for a source distance of 
10 Mpc and the energy being carried away by the gravitational waves ranging between  
$0.3\times 10^{51}$ and $2.8\times 10^{51}\ {\rm ergs}$. 
Second, we determine the types and frequencies of the fluid oscillation modes 
excited by the process. In particular, we find that the gravitational wave signals 
produced by the collapses are dominated by the fundamental quadrupole and quasi-radial modes
of the final equilibrium configurations. 
The two types of modes have comparable amplitude, with the latter mode representing the coupling
between the rotation and radial oscillations induced by the collapse.
In some collapse scenarios, we find that the oscillations are damped out within a few dynamical 
timescales due to the growth of differential rotations and the formation of strong shock waves.  
Third, we show that the spectrum of the gravitational wave signals is sensitive to the EOS,
implying that the detection of such gravitational waves could provide useful constraints 
on the EOS of newly born quark stars. 
Finally, for the range of rotation periods studied, we find no sign of the development of 
nonaxisymmetric dynamical instabilities in the collapse process.

\end{abstract}


\section{Introduction}
\label{sec:intro}

When nuclear matter is squeezed to a sufficiently high density, it turns
into uniform two-flavor quark matter ($u$ and $d$) and 
subsequently to three-flavor ($u$, $d$, and $s$) 
strange quark matter (SQM), since it is expected that SQM may be more 
stable than nuclear matter \citep{bod71,wit84}. 
Up to now there is still no definitive conclusion on the exact 
conditions for the phase transition to occur. 
But it is reasonable to assume that deconfined quark matter 
appears when the density is so high that the nucleons are 
``touching'' each other. 
At this point, where the number density of nucleons 
$n\sim 0(1\ {\rm fm}^{-3})$, the quarks lose their correlation 
with individual nucleons and appear in a deconfined phase.  
Since the density required for this to happen is not much higher than nuclear matter
density ($0.16\ {\rm fm}^{-3}$), the dense cores of neutron stars are 
the most likely places where the phase transition to quark matter may
occur astrophysically. 
It should be noted that strange quarks (in a confined phase) 
could already exist in neutron stars with a hyperon core. 

If SQM is absolutely stable, then once an SQM seed
is formed inside the core of a neutron star, 
it will begin to swallow the nuclear matter in its surroundings. 
The SQM region will grow, and
the neutron star will become a strange star consisting only of 
deconfined quarks. 
In principle, the existence of a thin crust of normal 
matter is possible at the surface of a strange star. 
On the other hand, if SQM is metastable at zero pressure, 
so that it is relatively more stable than nuclear matter only because 
of the high pressure in the cores of neutron stars, then the final 
products of the phase transition would be hybrid stars, which consist
of quark matter cores surrounded by normal matter outer parts.
In this case, the quark matter cores can be either in a pure quark 
matter phase or a mixed phase of quark and hadronic matter
(see, e.g., \citet{gle02,mad99} for reviews).   
In this paper, we do not distinguish between the two different 
scenarios, and we simply call the final product of the phase transition 
a quark star.
 
Since it was first realized that quark stars may exist in nature, 
much efforts has been put into their study. 
For example, the possibility that the conversion of neutron stars into 
strange stars might power cosmological $\gamma$-ray bursts has 
been considered (see, e.g., \citet{cheng96,ma96,bom00}).
Because the equation of state (EOS) describing the SQM is 
``softer'' than that of the nuclear matter,  
quarks stars are in general more compact than neutron stars.
The phase transition softens the EOS and leads to 
a collapse, and the conversion process liberates a huge amount 
of binding energy, on the order of $10^{52}$ ergs, which is required to 
power a $\gamma$-ray burst. 
Recently, the formation of quark stars has been invoked to explain 
the apparent bi modality of Type Ic supernovae \citep{pac05}.  
The prospect of detecting the gravitational wave signals emitted by quark 
stars has also been considered 
(e.g., \citet{cheng98,and02,mar02,gon03,oec04,lim05,yas05}).

On the observational side, it has recently been suggested that 
the compact X-ray source RXJ1856 is a quark star \citep{dra02}.
But the evidence is not yet conclusive. In fact, recent results
suggest that RXJ1856 is more likely to be a neutron star 
endowed with a strong magnetic field \citep{tho04}. 
The observational difficulty in distinguishing quark stars 
(if they exist) from neutron stars arises from the fact that gravity 
dominates over the strong interaction for compact objects with 
masses larger than $\sim 1.4M_{\odot}$. This in turn implies that there 
is no significant difference between the radii of neutron stars 
and quark stars.

One possible situation in which such phase-transition-induced collapses can occur 
is accreting neutron stars in low-mass $X$-ray binaries (LMXBs).
Neutron stars in these systems are spun up by 
accreting matter from their lower mass companions.
The neutron stars can accrete a mass $\gtrsim 0.5 M_{\odot}$ in a 
timescale of $\sim 10^8$ yr and become relatively massive
millisecond pulsars. The density in the cores of these neutron stars 
may be high enough for the nuclear matter to undergo phase transition.
It has been suggested \citep{cheng98} that the gravitational wave 
signals emitted from the phase-transition induced collapse of neutron stars
in these systems may be detectable by advanced gravitational wave detectors 
such as LIGO II.

While an initial estimate by \citet{oli87} suggested that the 
conversion of neutron stars into quark stars should proceed in a slow mode, 
later investigations by \citet{hor88} and \citet{lug94} pointed out that 
the conversion may in fact proceed at a much faster pace in a detonation mode, with the 
process being completed in a timescale about 0.1 ms. 
Nevertheless, it is important to stress that at the moment 
there is still no definitive conclusion on the timescale of this conversion 
process.

In this paper, we take the first step to investigate the dynamics 
of phase-transition-induced collapses 
by performing three-dimensional numerical simulations in Newtonian hydrodynamics and 
gravity. In particular, we study the gravitational wave 
signals emitted during the collapse of rotating neutron stars. 
We assume that (1) initially an SQM seed is formed inside the 
core of a neutron star, (2) the conversion process is completed 
within a timescale much shorter than the dynamical timescale of the star,
and (3) the result of the conversion is a quark star with a mixed phase 
of nuclear and quark matter inside the high density core 
surrounded by a normal nuclear matter region that extends to the star's 
surface.

Our initial stellar model before the phase transition
is a uniformly rotating equilibrium neutron star modeled by a polytropic 
EOS $P=k\rho^{\Gamma_0}$.
The collapse is induced by changing the polytropic 
EOS to a ``softer'' EOS for the mixed phase of quark and nuclear matter 
throughout the star initially. 
The star undergoes a gravitational collapse due to the reduction of the 
pressure. When the core of the star reaches a high enough density,  
the pressure gradient becomes large enough to balance the gravitational force and
the star bounces back and oscillates. 
Due to the coupling between the star's rotation and  
oscillations, the quadrupole mass moment of the star changes rapidly leading
to strong gravitational radiations.

The main focus of this paper is to study the features of the 
gravitational wave signals emitted from this kind of phase-transition-induced 
collapse of neutron stars. 
Our simulations suggest that the gravitational wave signals are dominated 
by the fundamental quasi-radial and quadrupole modes of the final 
equilibrium stars.  
While the frequencies of the fluid modes depend sensitively on the EOS 
model of the resulting quark stars, which is not yet established,
we find that the excitation of these two fluid modes is quite 
general and does not depend much on the EOS model. The detection of such 
gravitational wave signals would provide important information about the 
EOS of quark matter.
For the range of rotation periods we studied, we found no sign of 
development of nonaxisymmetric dynamical instabilities in the 
collapse process.  
For the collapse models considered in this paper, we find that the 
gravitational wave signals have frequencies of $\sim 2-3$ kHz 
and amplitudes of $\sim 10^{-22}$ for sources located 10 Mpc away,
and the energy carried away by the gravitational waves ranges from 
$0.27\times 10^{51}$ to $2.76\times 10^{51}\ {\rm ergs}$.

The plan of this paper is as follows:
In Sec.~\ref{sec:newton} we outline the Newtonian hydrodynamics 
equations and the finite differencing scheme used in the numerical simulations.
We also describe the numerical quadrupole formalism for the extraction of 
gravitational wave signals. 
Sec.~\ref{sec:EOS} describes the EOS models used in this paper. 
In Sec.~\ref{sec:collapse_result} we present the numerical results. 
Finally, we summarize and discuss our results in Sec.~\ref{sec:conclusion}.

\section{Newtonian Hydrodynamics Equations}
\label{sec:newton}

In this section, we present the system of Newtonian hydrodynamics 
equations for a non-viscous fluid flow and the numerical method that 
we use in the simulations. We also describe the numerical quadrupole 
formalism for the extraction of gravitational wave signals. 
Various consistency and convergence tests have been done to validate 
our numerical code.
This includes the ``standard'' Riemann shock tubes, static and boosted 
neutron stars, and binary neutron stars in orbit \citep{gre99}.
Recently, we have also demonstrated the capability of our code to 
evolve a rapidly rotating neutron star accurately for long-term simulation 
\citep{gre02,lin04}.

\subsection{Equations}
The system of equations describing the non viscous Newtonian 
fluid flow is given by 
\begin{equation}
        \frac{\partial \rho}{\partial t} + \nabla \cdot 
        \left( \rho {\bf v} \right) = 0 , 
\label{eq:rhoeq}
\end{equation}
\begin{equation}
        \frac{\partial}{\partial t} \left( \rho v_{i} \right) +
        \nabla \cdot \left( \rho v_{i} {\bf v} \right) + 
        \frac{\partial P}{\partial x_{i}} = - \rho 
        \frac{\partial \Phi}{\partial x_{i}} , 
\label{eq:momeq}
\end{equation}
\begin{equation}
        \frac{\partial \tau}{\partial t} 
        + \nabla \cdot \left(
        \left( \tau + P \right)
        {\bf v} \right) = -\rho {\bf v} \cdot \nabla \Phi  , 
\label{eq:taueq}
\end{equation}
where $\rho$ is the mass density of the fluid, ${\bf v}$ is the velocity
with Cartesian components $v_i$ ($i=1,2,3$), 
$P$ is the fluid pressure, $\Phi$ is the Newtonian
potential, and $\tau$ is the total energy density, defined by
\begin{equation}
        \tau = \rho \epsilon + \frac{1}{2} \rho {\bf v}^{2}  ,
\label{eq:tau_eps}
\end{equation}
in which $\epsilon$ is the internal energy per unit mass of the fluid.
The Newtonian potential $\Phi$ is determined by
\begin{equation}
\label{eq:potential}
        {\nabla}^{2} \Phi = 4 \pi G \rho  .
\end{equation}
The system is completed by specifying an EOS $P=P(\rho, \epsilon)$. 

The system of equations (\ref{eq:rhoeq})-(\ref{eq:taueq}) represents a 
hyperbolic system of first-order, flux-conservative equations in the form of:
\begin{equation}
\label{hyperbolic}
        \frac{\partial}{\partial t}\vec{U} + \nabla \cdot \vec{F} = \vec{S}.
\end{equation}
The evolved variables $\vec{U}$, the fluxes 
$\vec{F^j}$ and the sources $\vec{S}$ are given by
\begin{equation}
        \vec{U} = \left[
        \begin{array}{c}
              \rho \\
              \rho v_{i} \\
              \rho \epsilon + \frac{1}{2} \rho {\bf v}^{2}
        \end{array} \right] ,
\end{equation}
\begin{equation}
        \vec{{F}^{j}} = \left[
        \begin{array}{c}
                \rho v_{j} \\
                \rho v_{i} v_{j} + P\delta^{j}_{i}\\
                \left( \rho \epsilon + \frac{1}{2} \rho {\bf v}^{2} + P
                 \right)v_{j}
        \end{array} \right] ,
\end{equation}
\begin{equation}
        \vec{S} = \left[
        \begin{array}{c}
                0 \\
                - \rho \frac{\partial \Phi}{\partial x_i} \\
                - \rho {\bf v} \cdot \nabla \Phi
        \end{array} \right] .
\end{equation}
In each vector above, $i$ ranges from 1 to 3 (for $x$, $y$, and $z$), 
thus making the vector five-dimensional.  Also note
that in the case of $\vec{F^j}$, $j$ goes from 1 to 3, so that
we have a total of five vectors, each with five components.

\subsection{Spectral Decomposition of the Characteristic Matrix}

Our Newtonian hydrodynamics code uses a high resolution shock
capturing (HRSC) scheme \citep{hir92} to solve Eq.~(\ref{hyperbolic}).
A HRSC scheme, using either exact or approximate Riemann solvers, with 
the characteristic fields (eigenvalues) of the system, has the 
ability to resolve discontinuities in the solution (e.g., shock waves).
Moreover, such schemes have high accuracy in regions where the fluid flow
is smooth. 

To implement the HRSC schemes, it is necessary to decompose the
characteristic matrix of the system, $\frac{\partial
\vec{F^{j}}}{\partial \vec{U}}$, to determine the characteristic
fields (eigenvectors) and the characteristic speeds of those fields
(the corresponding eigenvalues).  For any EOS in which
the pressure $P$ is a function of $\rho$ and $\epsilon$, the speed of
sound in the fluid $c_s$ is given by
\begin{equation}
        c_s^2 = {\left. \frac{\partial P}{\partial \rho} \right|}_{\cal S}
        = {\left. \frac{\partial P}{\partial \rho} \right|}_{\epsilon} + 
        \frac{P}{\rho ^{2}}{\left . \frac{\partial P}{\partial \epsilon}
        \right|}_{\rho} ,
\end{equation}
where ${\cal S}$ is the entropy per particle.  The characteristic matrix
possesses a complete set of linearly independent eigenvectors ${\vec{r}}_i$
rendering the system strongly hyperbolic.  
For example, the $x$ component of the flux term, $\vec{F^x}$, 
in Eq.~(\ref{hyperbolic}), has the characteristic matrix 
$\frac{\partial \vec{F^{x}}}{\partial \vec{U}}$.  Its
five eigenvectors satisfy the equation
\begin{equation}
        \left[ \frac{\partial \vec{F^{x}}}{\partial \vec{U}} \right] 
        \left[ {\vec{r}}_i \right] = {\lambda}_i 
        \left[ {\vec{r}}_i \right]  \ ; \ i=1,\cdots,5,
\end{equation}
where the eigenvalues $\lambda_i$ are given by:
\begin{equation}
\label{eigenvalues}
        \lambda_1 = \lambda_2 = \lambda_3 = v_x  ,
\end{equation}
\begin{equation}
        \lambda_4 = v_x + c_s  ,
\end{equation}
\begin{equation}
        \lambda_5 = v_x - c_s  .
\end{equation}
Explicitly, the five right eigenvectors of the Jacobian are
\begin{equation}
        \vec{r}_1 = {\left[ 1, v_x, v_y, v_z, H-\frac{c_s^2}{\gamma - 1} 
        \right]}^T  ,
\end{equation}
\begin{equation}
        \vec{r}_2 = {\left[ 0, 0, \rho, 0, \rho v_y \right]}^T  ,
\end{equation}
\begin{equation}
        \vec{r}_3 = {\left[ 0, 0, 0, \rho, \rho v_z \right]}^T  ,
\end{equation}
\begin{equation}
        \vec{r}_4 = \frac{\rho}{2 c_s} {\left[ 1, v_x + c_s, v_y, v_z, 
        H + v_x c_s \right]}^T  ,
\end{equation} 
\begin{equation}
\label{eigenvectors}
        \vec{r}_5 = \frac{\rho}{2 c_s} {\left[ 1, v_x - c_s, v_y, v_z, 
        H - v_x c_s \right]}^T  ,
\end{equation}
where the total enthalpy $H = \epsilon + \frac{{\bf v}^2}{2} +
\frac{P}{\rho}$ is introduced along with $\gamma = 1 + \frac{1}{\rho}
\frac{\partial P}{\partial \epsilon}$. The superscript $T$ denotes
transpose.  The eigenvalues and eigenvectors for the $y$ and $z$
components can be obtained from a cyclic permutation of spatial
indices and components of each $\vec{r}_i$.

\subsection{Finite Differencing Method}

To solve equations (\ref{eq:rhoeq})-(\ref{eq:taueq})
numerically, the system must be suitably discretized in both space and
time.  
In our code we use a discretization scheme that is second-order
accurate in time. The spatial order, however, depends on which 
cell reconstruction method we choose (see below). 

We first consider the discretization in space.  We show the
discretization in the $x$-direction only; the other spatial
dimensions are handled in a similar manner.  We also consider the
equations in the absence of source terms since the discretization of the
source terms does not present any difficulties.

The problem, then, is reduced to finding the appropriate update of the state
vector $\vec{U}$ for fluxes in the $x$-direction:
\begin{equation}
\frac{\partial \vec{U}}{\partial t} + \frac{\partial \vec{F^x}}{\partial x}
= 0 .
\end{equation}
The spatial discretization of this equation can be written in the standard
manner:
\begin{equation}
\frac{\partial \vec{U}_i}{\partial t} + 
\frac{(\vec{f^{\ast}})_{i+1/2} - (\vec{f^{\ast}})_{i-1/2}}{\Delta x} = 0 ,
\end{equation}
where $(\vec{f^{\ast}})_{i \pm 1/2}$ is the ``numerical flux'' evaluated at
the interfaces $i \pm 1/2$ of the spatial cell $i$ using information from 
the left and right sides. The state variables $\rho$, $\epsilon$, and
${\bf v}$, which are cell-centered quantities, must be properly
reconstructed and interpolated to the cell interfaces. 
This cell-reconstruction procedure controls the spatial order of the 
numerical scheme. In our implementation, we use the third-order 
piecewise parabolic method (PPM \citet{col84}) for the reconstruction. 

To calculate the numerical fluxes, we make use of Roe's approximate Riemann
solver \citep{roe81}. Roe's solver is well-established and widely used.  
It has good shock capturing capability and can be used for a wide range
of equations of state.  For a detailed discussion of
the Roe's scheme, see for example, \citet{hir92}.
With the fluid state interpolated to the cell interfaces, the
formula for the numerical flux then becomes
\begin{equation}
(\vec{f^{\ast}})_{i+1/2} = \frac{1}{2} \left[  \vec{F^{x}_L} + \vec{F^{x}_R}
- \sum_{j=1}^5 | \tilde{\lambda}_j | {\Delta \tilde{\omega}_j} 
\tilde{\vec{r}_j} \right] .
\end{equation}
Here $\vec{F^{x}_L}$ and $\vec{F^{x}_R}$ are the fluxes computed using,
respectively, the primitive variables to the left and right of the interface
between the $i$-th and $(i+1)$-th cells.  Both $\tilde{\lambda}_j$ and 
$\tilde{\vec{r}_j}$ are computed by 
Eqs.~(\ref{eigenvalues})-(\ref{eigenvectors}) 
using the following weighted quantities:
\begin{equation}
\tilde{\rho} = \sqrt{\rho_i \rho_{i+1}} ,
\end{equation}
\begin{equation}
\tilde{\bf v} = \frac{ ({\bf v} \sqrt{\rho})_{i} + 
({\bf v} \sqrt{\rho})_{i+1}}{ \sqrt{\rho_i} + \sqrt{\rho_{i+1}}}  ,
\end{equation}
\begin{equation}
\tilde{H} = \frac{ (H \sqrt{\rho})_{i} + 
(H \sqrt{\rho})_{i+1}}{ \sqrt{\rho_i} + \sqrt{\rho_{i+1}}} .
\end{equation}
Finally, the quantities $\Delta \tilde{\omega}_j$ represent the jumps in the
characteristic variables across each characteristic field, and are obtained
from 
\begin{equation}
\vec{U}_R - \vec{U}_L = \sum_{j} {\Delta \tilde{\omega}_j} 
\tilde{\vec{r}_j} .
\end{equation}

Next we turn to temporal discretization. To achieve second-order accuracy
in time, we employ a basic two-step method.  Given the fluid state at the mesh
points $\vec{U}_i^n$, where $i$ represents the spatial mesh and $n$ represents
the time coordinate of the state variables, we calculate $\vec{U}_i^{n+1/2}$,
the fluid state midway between the time $t$ and $t + \Delta t$, to first order:
\begin{equation}
\label{firstorder}
\frac{\vec{U}_i^{n+1/2}-\vec{U}_i^n}{(\Delta t/2)} + 
\frac{(\vec{f^{\ast}})_{i+1/2}^n - (\vec{f^{\ast}})_{i-1/2}^n}{\Delta x} = 
\vec{S}_i^n .
\end{equation}
Here we have included the source term $\vec{S}_i^n$.  
The fluid state $\vec{U}_i^{n+1}$ is then given by:
\begin{equation}
\label{secondorder}
\frac{\vec{U}_i^{n+1}-\vec{U}_i^n}{\Delta t} + 
\frac{(\vec{f^{\ast}})_{i+1/2}^{n+1/2} - 
(\vec{f^{\ast}})_{i-1/2}^{n+1/2}}{\Delta x} = \vec{S}_i^{n+1/2}  .
\end{equation}

\subsection{Generation of gravitational waves}
\label{sec:G_wave}

We calculate the gravitational wave signals emitted during a gravitational 
collapse using the standard quadrupole approximation, 
which is valid for a nearly Newtonian system (see, e.g., the recent review by 
\citet{fla05} for details).
The power emitted in gravitational waves is given by
\begin{equation} 
{dE\over dt}={G\over 5c^5} {Q}_{ij}^{(3)} {Q}_{ij}^{(3)} ,
\label{eq:GW_dEdt}
\end{equation}
where the superscript (3) indicates the third time derivative and 
the traceless quadrupole moment of the mass distribution $Q_{ij}$ is 
defined by 
\begin{equation}
Q_{ij}=\int{ \rho (x^i x^j - {1\over 3}\delta^{ij}r^2) d^3x} .
\end{equation}
The total energy radiated in the form of gravitational wave is 
calculated by 
\begin{equation}
\Delta E_{\rm GW}(t) = \int_0^t dt {dE\over dt} .
\end{equation}

For simplicity we evaluate the gravitational waveforms on the $x$-axis
(where the $z$-axis is the rotation axis). The nonvanishing 
components of the leading order radiative gravitational field $h_{ij}^{\rm TT}$
are 
\begin{equation}
h_+ \equiv h_{yy}^{\rm TT} = - h_{zz}^{\rm TT} 
     = {G\over c^4 r} \left( \ddot{Q}_{yy} - \ddot{Q}_{zz} \right) ,
\end{equation}

\begin{equation}
h_\times \equiv h_{yz}^{\rm TT} = {2G\over c^4 r} \ddot{Q}_{yz} .
\end{equation}
where $\ddot{Q}_{ij}$ denotes the second time derivative of $Q_{ij}$.
Using direct finite-differencing method to calculate
the high-order time derivatives of $Q_{ij}$ in numerical simulation would 
generate a large amount of unwanted noise.
In our simulations, we follow a standard approach to reexpress $\ddot{Q}_{ij}$ as 
an integral over the fluid variables and calculate them on each time slice 
(see, e.g., \citet{bla90,fin90,zwe97}).

We also calculate the gravitational wave signal using the angle-averaged 
waveforms for the polarization states $h_{+}$ and $h_{\times}$ 
\citep{cen93}:
\begin{equation}
< (rh_{+})^2 > = {1\over 4\pi} \int (r h_{+})^2 d\Omega ,
\end{equation}

\begin{equation}
< (rh_{\times})^2 > = {1\over 4\pi} \int (r h_{\times})^2 d\Omega ,
\end{equation}
where $d\Omega$ is a solid angle element. The angle brackets denote 
averaging over all source orientations.
In terms of $\ddot{Q}_{ij}$ they are given explicitly by 
\citep{cen93}

\begin{eqnarray}
< (r h_{+})^2 > &=& {G\over c^8}\left[{4\over 15} ( \ddot{Q}_{xx} - 
\ddot{Q}_{zz} )^2 + {4\over 15} ( \ddot{Q}_{yy}-\ddot{Q}_{zz} )^2 
+ {1\over 10} ( \ddot{Q}_{xx}-\ddot{Q}_{yy} )^2 + 
{14\over 15} ( \ddot{Q}_{xy} )^2   \right. \cr
&& \cr
&&\left. 
+ {4\over 15} ( \ddot{Q}_{xz} )^2 + {4\over 15} (\ddot{Q}_{yz} )^2 \right] , 
\end{eqnarray}
\begin{equation}
< (r h_{\times})^2 > = {G\over c^8}\left[ {1\over 6} 
( \ddot{Q}_{xx} - \ddot{Q}_{yy} )^2 + {2\over 3}( \ddot{Q}_{xy} )^2 
+ {4\over 3} ( \ddot{Q}_{xz} )^2 + {4\over 3} ( \ddot{Q}_{yz} )^2 \right] .
\end{equation}

\section{Equation of State}
\label{sec:EOS}

\subsection{Initial neutron star model}
\label{sec:EOS_initNS}

We model our initial equilibrium rotating neutron star before the phase 
transition by a polytropic EOS 
\begin{equation}
P=k_0\rho^{\Gamma_0} ,
\label{eq:poly_EOS}
\end{equation}
where $k_0$ and $\Gamma_0$ are constants. 
On the initial time slice, we also need to specify the specific internal 
energy $\epsilon$. For the polytropic EOS, the thermodynamically consistent 
$\epsilon$ is given by 
\begin{equation}
\epsilon = {k_0\over \Gamma_0-1} \rho^{\Gamma_0-1} .
\end{equation}
Note that the pressure in Eq.~(\ref{eq:poly_EOS}) can also be written as
\begin{equation} 
P=(\Gamma_0-1)\rho\epsilon .
\end{equation}

\subsection{Quark matter}

In the literature, the so-called MIT bag model EOS has been used
extensively to describe quark matter inside compact stars 
(see, e.g., \citet{gle02,mad99}). 
In our study, we assume that the quarks are massless. 
We note that while it is a good approximation to neglect the $u$- and
$d$-quark masses, since they are much smaller than the quark chemical 
potential, which is on the order of 300 MeV, this is not the case for 
the $s$-quark mass, $m_s\sim O(100)$ MeV. Nevertheless, the inclusion 
of the $s$-quark mass would decrease the pressure by only a few 
percent \citep{alc88}. Hence, we do not expect that the inclusion of the 
$s$-quark mass would change our results qualitatively.

Assuming that the quarks are massless, the MIT bag model EOS is 
given by
\begin{equation}
P = {1\over 3}(\rho_{\rm tol} - 4 B) ,
\label{eq:mit_eos}
\end{equation}
where $\rho_{\rm tol}$ is the (rest frame) total energy density 
and $B$ is the so-called bag constant.
The quarks are asymptotically free in the limit 
$\rho_{\rm tol}\rightarrow \infty$. At finite densities, the 
quark interactions, and hence the confinement effect must be taken 
into account. The bag constant $B$ is a phenomenological parameter 
that describes this confinement. 
It should be noted that at the moment there is no general consensus 
on the best value of $B$ to describe quark matter.
The result from fits of the hadronic spectrum gives 
$B^{1/4} \approx 145$ MeV (in units of $\hbar=c=1$) \citep{deg75}, 
while the calculation of hadronic structure functions 
suggests $B^{1/4} \sim 170$ MeV \citep{ste95}. 
On the other hand, lattice QCD results suggest a value up to 
$B^{1/4}\sim 190$ MeV \citep{satz82}. 
Many investigators have used different values of $B$ within 
the range $B^{1/4}\sim 145-200$ MeV to study quark matter inside neutron 
stars. 
In our simulations, we employ the MIT bag model with 
$B^{1/4}=170$ MeV for the quark matter in the mixed phase EOS 
described below.

In our Newtonian simulations, we use the rest-mass density $\rho$ 
and specific internal energy $\epsilon$ as fundamental variables in
the hydrodynamics equations. The total energy density $\rho_{\rm 
tol}$, which includes the rest-mass contribution, is decomposed as 
(in our code units $G=c=M_{\odot}=1$)
\begin{equation}
\rho_{\rm tol} = \rho + \rho\epsilon . 
\label{eq:rho_tol_define}
\end{equation}
In terms of our hydrodynamic variables the expression for the 
MIT bag model becomes 
\begin{equation}
P_{\rm q} = {1\over 3} \left( \rho + \rho\epsilon - 4B\right) .
\label{eq:quark_eos}
\end{equation}
The MIT bag model is a relativistic EOS. One might worry that using this EOS 
in a Newtonian simulation is somewhat inconsistent. 
However, it should be noted that the MIT bag model is a microscopic 
phenomenological model, whereas the Newtonian hydrodynamic equations are 
used to describe the global structure of the star.

\subsection{Mixed Phase}
\label{sec:Eos_mixed}

As discussed in the introduction, we assume that once a SQM seed 
is formed inside the core of a neutron star, the star will transform to a 
quark star on a timescale much shorter than the dynamical timescale of the 
neutron star. In this short timescale limit, we do not need to model the 
complicated and poorly known dynamical process of the conversion itself. 
Instead, we focus on the collapse process of the star 
due to the softening of the EOS after the conversion is completed.
In our simulations, the collapse is induced by changing the initial 
polytropic EOS (see Eq. (\ref{eq:poly_EOS})) to a ``softer'' EOS to 
describe the quark star in the initial time slice.

The model EOS that we consider for the quark star is composed of two parts:
(1) a mixed phase of quark and nuclear matter in the core at a density higher 
than a certain critical value $\rho_{\rm NM}$; (2) a normal nuclear matter 
region extending from $\rho=\rho_{\rm NM}$ to the surface of the star. 
Explicitly, the pressure is given by 
\begin{equation}
P = \left\{ \begin{array}{cc}
           \alpha P_{\rm q} + (1-\alpha) P_{\rm n}
         & \ \mbox{for} \ \  \rho > \rho_{\rm NM} \\
        \\
            P_{\rm n} 
         & \mbox{for} \ \ \rho  \leq \rho_{\rm NM} ,  \end{array}  \right.
\label{eq:mixed_EOS}
\end{equation}
where    
\begin{equation}
\alpha = \left\{ \begin{array}{cc}
          { (\rho -\rho_{\rm NM}) / (\rho_{\rm QM} - \rho_{\rm NM}) } 
         & \ \mbox{for} \ \  \rho_{\rm NM} < \rho < \rho_{\rm QM} \\
        \\
            1 
         & \mbox{for} \ \ \rho_{\rm QM} < \rho ,  \end{array}  \right.
\label{eq:alpha}
\end{equation}
is defined to be the scale factor of the mixed phase, and where
$P_q$ is the pressure contribution of the quark matter, while $P_n$ 
is that of the nuclear matter.
In Eq.~(\ref{eq:alpha}), $\rho_{\rm QM}$ is defined to be the critical 
density above which a pure quark matter phase is energetically favored
over the mixed phase.

We take $P_q$ to be given by the MIT bag model (see Eq.~(\ref{eq:quark_eos}))
with a bag constant $B^{1/4}=170$ MeV. 
The critical density of the phase transition from nuclear matter to pure 
quark matter is model dependent; it could range from 4 to $8 \rho_{\rm nuc}$ 
\citep{cheng98,hae03,bom04}, where 
$\rho_{\rm nuc}=2.8\times 10^{14}\ {\rm g\ cm^{-3}}$ 
is the nuclear density. \citet{cheng96} have argued that the accretion 
induced phase transition from neutron stars to strange stars in low-mass 
X-ray binaries could be the origin of $\gamma$-ray bursts. They choose the transition density 
$\sim 8-9 \rho_{\rm nuc}$, so that the phase transition cannot occur until 
a large amount of mass has been transferred from the lower mass companion 
to the neutron star; otherwise, there 
may be too many $\gamma$-ray bursts produced from this mechanism. 
We choose $\rho_{\rm QM}=9\rho_{\rm nuc}$ in our simulation, 
which can produce stronger gravitational radiation.  
Other choices of $\rho_{\rm QM}$ are not ruled out but 
the gravitational radiation will be relatively weaker.
In all cases studied in this paper, the maximum value of $\alpha$ is 
always less than unity during the time evolutions.
Also, we define the transition density from the mixed phase to nuclear 
matter $\rho_{\rm NM}$ to be at the point where $P_q$ vanishes initially.
This corresponds to $\rho_{\rm NM}= 7.25\times 10^{14}\ 
{\rm g\ cm^{-3}} = 2.6 \rho_{\rm nuc}$ for $B^{1/4}=170$ MeV in our study.

For the normal nuclear matter, we use an ideal-fluid model 
\begin{equation}
P_{\rm n}=(\Gamma-1)\rho \epsilon ,
\label{eq:eos_idealgas}
\end{equation}
where $\Gamma$ is the effective adiabatic index. 
In our mixed-phase EOS model, we allow for the possibility that the nuclear 
matter EOS can be different from that before the phase transition. This is 
achieved by using values of $\Gamma$ different from the initial value 
$\Gamma_0$. In particular, we choose $\Gamma < \Gamma_0$ in our simulations.
It is possible that the nuclear matter may not be stable during the phase 
transition process, and hence some quark seeds could appear inside 
the nuclear matter. In the presence of the quark seeds in the nuclear matter, 
the effective adiabatic index will be reduced.
In our simulations, $B$, $\rho_{\rm NM}$, and $\rho_{\rm QM}$ are fixed to be 
the values given above for all cases studied. We consider $\Gamma$ as 
the only free parameter of the mixed phase EOS.

\section{Numerical Results}
\label{sec:collapse_result}

\subsection{General Considerations}
\label{sec:mode_identity}

The gravitational collapse of compact stellar objects is one of the most promising 
sources of gravitational waves for interferometer detectors such as LIGO, VIRGO,
GEO600, and TAMA300. 
In the past decade, numerous multidimensional (two- or three-dimensional) hydrodynamic simulations 
have been performed to study the gravitational wave signals emitted from some 
collapse processes. 
For example, the signals from the rotational core collapse to a proto neutron star 
have been studied in both Newtonian (e.g., \citet{fin90,mon91,bon93,zwe97,ram98,fry02}) 
and general relativistic frameworks \citep{dim02,sie03,shi04,dim05,shi05}.
While it is still consistent to use Newtonian dynamics to study the rotational core
collapse, as gravity is relatively weak in this process, fully general relativistic 
simulation is required to study the more extreme scenario: the gravitational 
collapse of rotating neutron stars to black holes. The gravitational wave signals
emitted from such a process were studied 20 years ago by \citet{sta85} in two dimensions, 
and more recently by \citet{bai05} in three dimensions. 

In this paper, we study the gravitational wave signals emitted from the 
gravitational collapses of rapidly rotating neutron stars to quark stars. 
In this collapse scenario, gravity is stronger than that in the rotational core 
collapse, but still not strong enough to produce a black hole 
(we will return to this point later). To the best of our knowledge, our work 
is the first hydrodynamic simulation to study this kind of collapse process and 
the emitted gravitational wave signals in three dimensions.

Before we present our numerical results for the phase-transition induced 
collapse models based on the mixed phase EOS discussed in Sec.~\ref{sec:Eos_mixed}, in this 
section we study the features of the gravitational wave signals emitted from such process 
using a simplified collapse model. 
Our simulations suggest that the emitted gravitational wave signals 
are dominated by two fluid oscillation modes. 
In particular, we demonstrate that these two modes are the fundamental 
quasi-radial and quadrupole modes of the final equilibrium star. 
The excitation of these fluid modes is a general feature that 
does not depend sensitively on the EOS model.

In this study, we use the same initial neutron star configuration as that of 
the reference model R in the mixed-phase collapse models 
(see Sec.~\ref{sec:num_models}).
The neutron star is modeled by a polytropic EOS $P=k \rho^2$, with the 
polytropic constant $k=k_0=60$. 
The collapse is induced by reducing the pressure by using a smaller 
polytropic constant $k=54$ (i.e., a 10\% reduction of the pressure)
during the evolution.
This simplified collapse model has the advantage that it can produce the 
general feature of the gravitational waveforms that we see 
in the mixed-phase model, while remaining simple enough that the fluid oscillation modes 
of the final equilibrium configuration can be studied in detail.

\begin{figure*}
  \begin{minipage}{0.4\linewidth}
        \centering
        \includegraphics*[width=6.5cm]{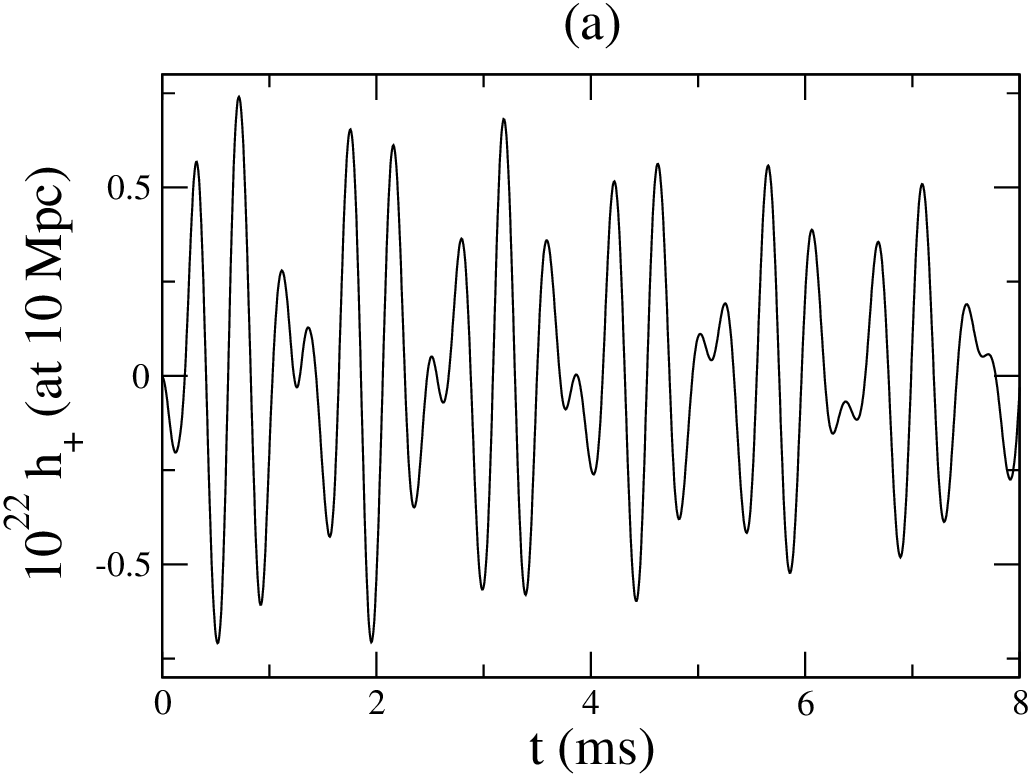}
  \end{minipage}%
  \begin{minipage}{0.4\linewidth}
        \centering
        \includegraphics*[width=5.5cm]{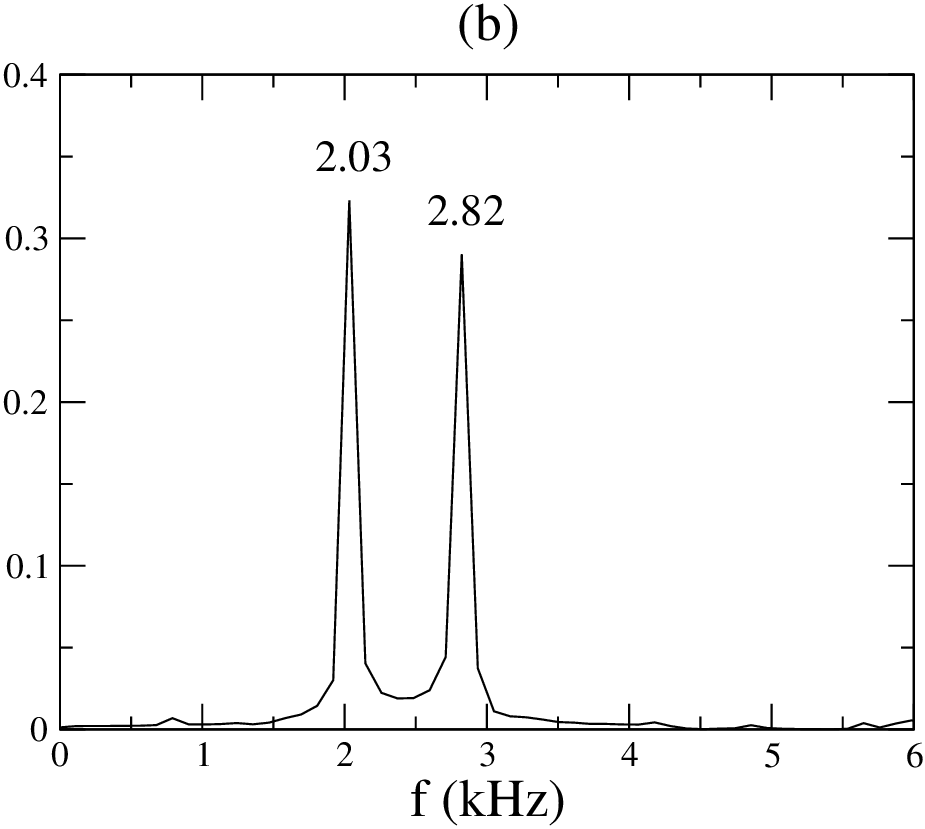}
  \end{minipage}
        \caption{(a) Gravitational waveform $h_+$ on the $x$ axis for a simplified 
	collapse model. The initial neutron star is the same as that of the reference model R
	in Table 1.	
	The neutron star is modeled by a polytropic EOS $P=k \rho^2$, with $k=60$. 
	The collapse is induced simply by reducing the value of $k$ by 10\%.
	(b) Fourier transform of $h_+$ for the same case.}     
        \label{fig:coll_byk_h+}
\end{figure*}

In Fig.~\ref{fig:coll_byk_h+} (a) we show the evolution of the gravitational 
wave amplitude $h_+$ on the $x$-axis for this collapse model. 
For all cases studied in this paper, it is assumed that the source is located 
at a distance of 10 Mpc (i.e., the distance to nearby galaxies). 
Our simulations are performed in full three dimensions without any symmetry
assumptions. Any non-axisymmetric fluid flow during the evolution can 
in general produce a nonzero contribution to the wave field $h_{\times}$.
Nevertheless, we see that for all models studied in this paper
$h_{\times}$ is essentially zero (within numerical accuracy) during the evolution. 
This suggests that the fluid flow remains axisymmetric during the 
collapse and no non axisymmetric fluid oscillations are excited. 
This implies that for the range of rotation periods studied, there are no 
nonaxisymmetric dynamical instabilities in the collapse process. 
Hence, we will consider only $h_+$ in the rest of this paper.

Fig.~\ref{fig:coll_byk_h+} (a) shows clearly that the waveform is not dominated by 
a single oscillation mode. 
The Fourier transform of $h_+$ plotted in Fig.~\ref{fig:coll_byk_h+} (b) 
shows that the waveform is composed of two oscillation modes at frequencies 
2.03 and 2.82 kHz. 
To show that these two excited modes are respectively the fundamental 
quadrupole and quasi-radial modes, we compare the mode frequencies with 
those obtained by perturbing a configuration that is the same as the final 
equilibrium star.

\begin{figure}
\includegraphics*[width=6.8cm]{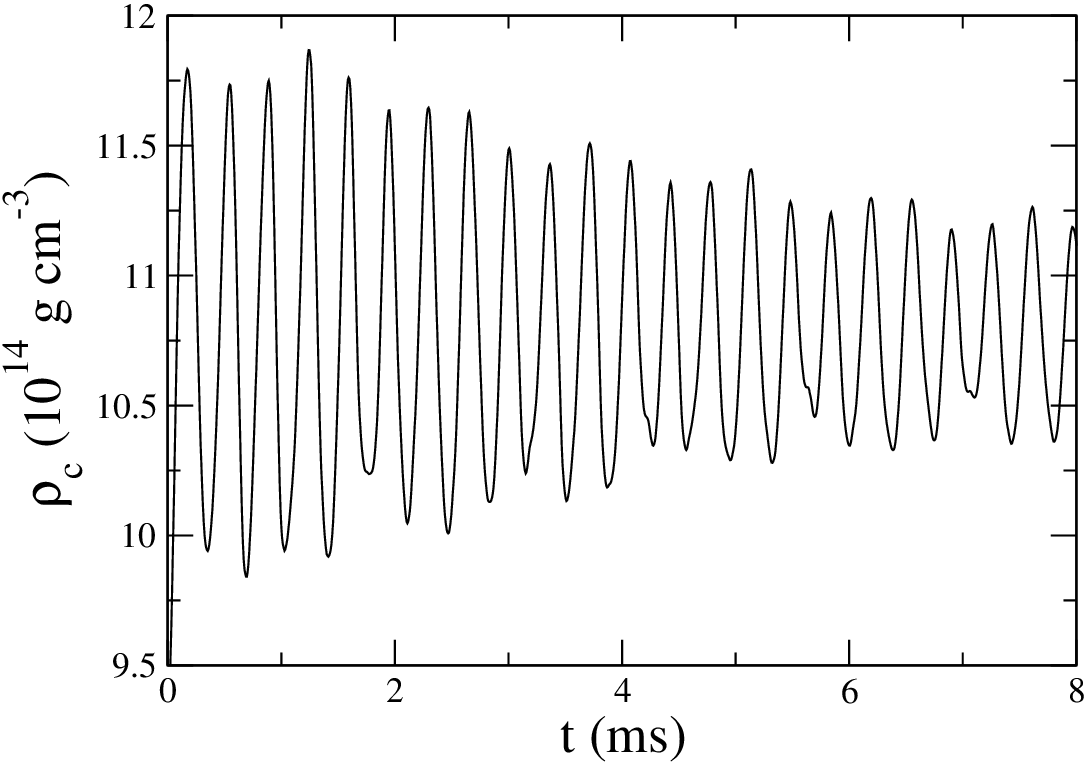}
\caption{Time evolution of the central density for the simplified collapse 
model.}  
\label{fig:coll_byk_rhoc}
\end{figure}

To begin, we first need to construct an unperturbed configuration that is 
the same as the final equilibrium star resulting from the 
collapse. In Fig.~\ref{fig:coll_byk_rhoc} we show the time evolution of the 
central density $\rho_{\rm c}$. 
It is seen that $\rho_{\rm c}$ oscillates around its final 
equilibrium value after the initial collapse phase. However, it is 
computationally too expensive to run the simulation until the final equilibrium 
state is reached. 
Instead, we use the time-average value of $\rho_{\rm c}$ in the period $2-6$ ms,
$1.08\times 10^{15}\ {\rm g\ cm^{-3}} $, 
to approximate the central density of the final equilibrium configuration. 
To fully specify the configuration, we also need to 
determine the final spin rate. More generally, we need the profile of the 
rotational velocity, since differential rotation can develop during the 
collapse. As we show in Sec.~\ref{sec:waveforms}, for some of the mixed-phase 
collapse models studied in this paper, the stellar oscillations triggered by the 
collapse are damped very quickly by the growth of differential rotation. 
However, it has been shown by \citet{ste04} that the frequencies of the 
quasi-radial and quadrupole modes do not depend sensitively on rotation 
along a sequence of stellar models with the same central density. 
Hence, for our purpose, it is accurate enough to study the oscillations of 
a nonrotating star model that has the same central density as the final 
equilibrium star produced by the rotational collapse. 

\begin{figure}
\includegraphics*[width=6.8cm]{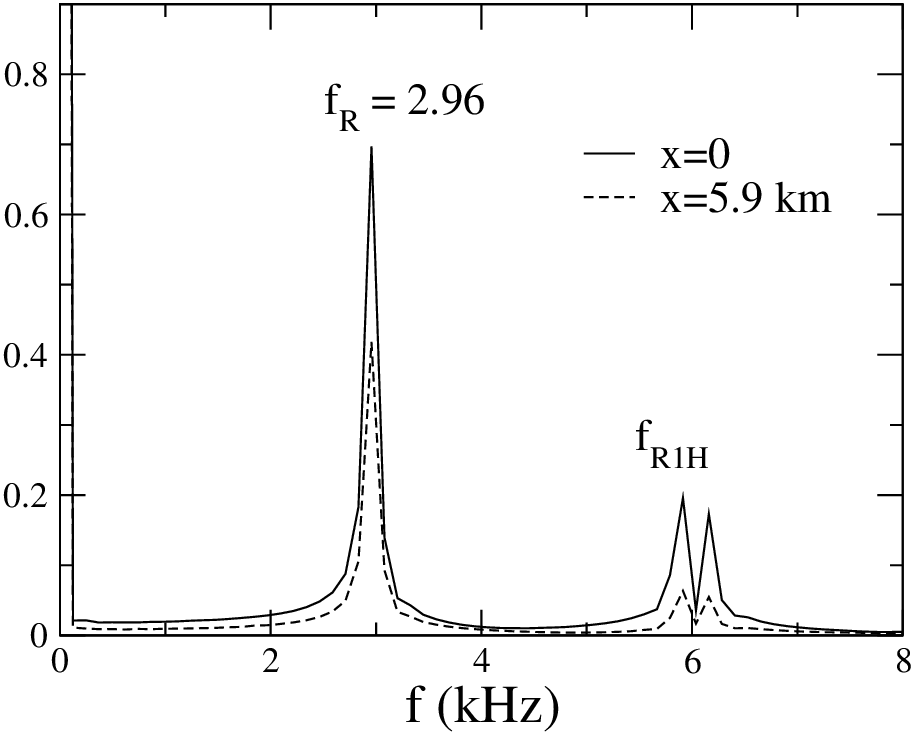}
\caption{Fourier transforms of the time evolution of the density at 
$x=0$ and 5.9 km for the radially perturbed nonrotating model discussed in 
the text.
The peaks at $f_{\rm R}$ and $f_{\rm R1H}$ correspond to the fundamental 
radial mode and its first harmonic respectively.  }
\label{fig:l0perturb_rhocFT}
\end{figure}

The unperturbed nonrotating star is modeled by the polytropic EOS with 
$k=54$. It has a central density 
$\rho_{\rm c}=1.08\times 10^{15}\ {\rm g\ cm^{-3}} $, and baryonic mass 
$M=1.73 M_{\odot}$ and radius $R=13.46$ km. 
To excite a particular mode we add initial perturbation to the equilibrium 
star. For the fundamental radial ($l=0$) mode, we  
perturb the velocity by adding a radial perturbation of the form
\begin{equation}
\delta v_r = \alpha \sin\left( { \pi r\over R } \right) ,
\end{equation}
where $\alpha$ is the amplitude of the perturbation. 
We use $\alpha=0.03$, which corresponds to 3\% of the speed of light. 
We follow the evolution of the perturbed star using our fully nonlinear 
evolution code. 
In Fig.~\ref{fig:l0perturb_rhocFT} we show the Fourier transforms of the 
time evolution of the density at $x=0$ and 5.9 km. 
Fig.~\ref{fig:l0perturb_rhocFT} shows that the fundamental radial mode at 
frequency $f_{\rm R}=2.96$ kHz is excited and dominates the time evolution. 
It is seen that the first harmonic $f_{\rm R1H}$ has also been excited.

For a nonrotating star, the radial mode does not radiate gravitational wave. 
However, the non radial modifications of the mode eigenfunction due to rotation
implies that the radial mode becomes quasi-radial and is capable of emitting 
gravitational waves \citep{chau67}.
We see that the frequency of the fundamental radial mode $f_{\rm R}=2.96$ kHz
agrees with the second peak in Fig.~\ref{fig:coll_byk_h+} (b) to about 5\%. 
This shows that the second peak in Fig.~\ref{fig:coll_byk_h+} (b), at a 
frequency of 2.82 kHz, corresponds to the fundamental quasi-radial mode of the 
final equilibrium star. 
The fact that the frequencies are so close to one another
reinforces the conclusion of \citet{ste04}, that the frequency of the 
fundamental quasi-radial mode does not depend sensitively on rotation
for a sequence of stellar models with fixed central density.

\begin{figure*}
  \begin{minipage}{0.4\linewidth}
        \centering
        \includegraphics*[width=5.5cm]{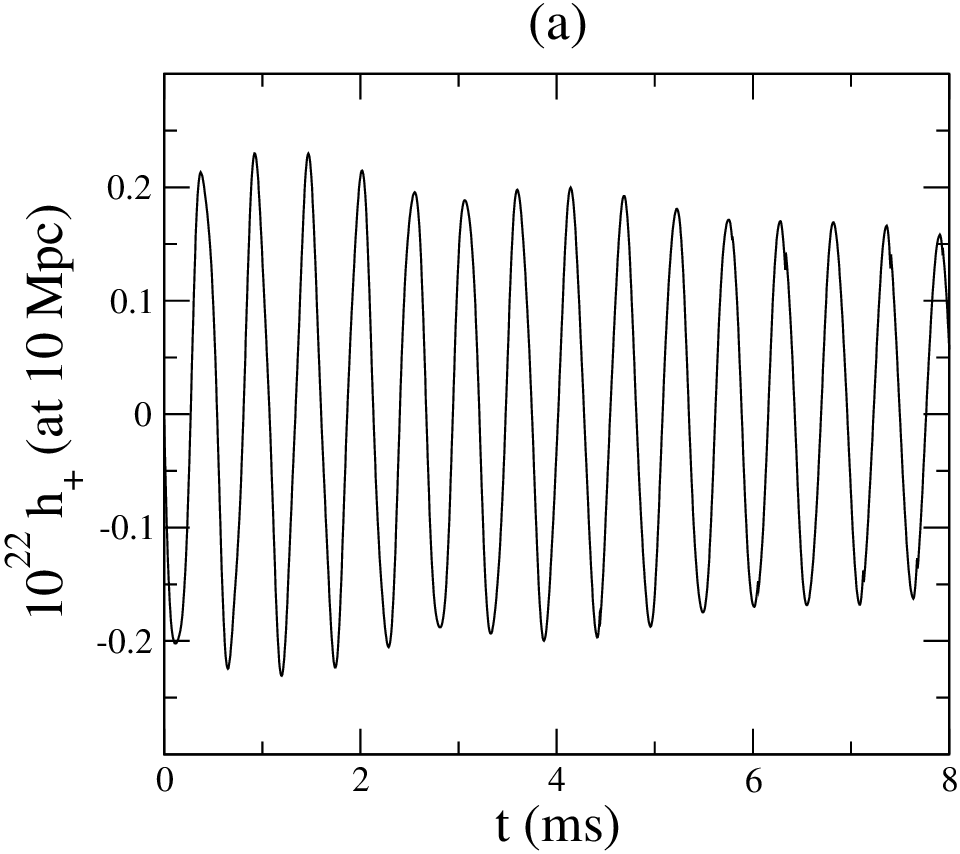}
  \end{minipage}%
  \begin{minipage}{0.4\linewidth}
        \centering
        \includegraphics*[width=5.0cm]{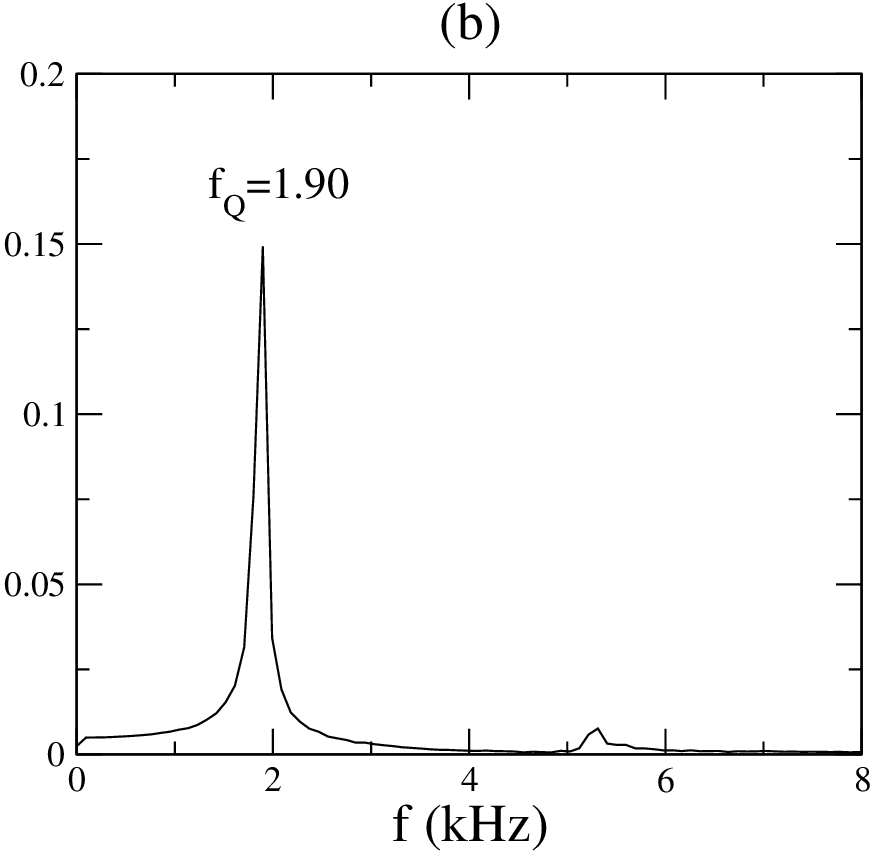}
  \end{minipage}
        \caption{(a) Time evolution of the gravitational wave amplitude $h_+$ 
          on the $x$-axis emitted by the quadrupole mode of the nonrotating 
          model discussed in the text. (b) Fourier transform of $h_+$ for the 
          same case.} 
        \label{fig:l2perturb}
\end{figure*}

To excite the fundamental quadrupole mode, we perturb the 
$\theta$-component of the velocity by \citep{font01,ste04} 
\begin{equation}
\delta v_{\theta} = \alpha \sin\left( {\pi r\over R} \right)
                    \sin\theta \cos\theta .
\end{equation}
The form of the polar-angle ($\theta$) dependence is motivated 
by the fact that the $\delta v_{\theta}$ of spheroidal 
oscillation modes is proportional to $\partial Y_{lm}/ \partial \theta$,
where $Y_{lm}$ represents the usual spherical harmonics and we want to excite the 
$l=2$, $m=0$ mode. 
In Fig.~\ref{fig:l2perturb} (a) we show the time evolution of the 
gravitational wave amplitude $h_+$ on the $x$ axis. 
The waveform is dominated only by the excited quadrupole mode. 
The Fourier transform of $h_+$ plotted in Fig.~\ref{fig:l2perturb} (b) 
shows that the frequency of the quadrupole mode is $f_{\rm Q}=1.90$ kHz, which 
agrees with the first peak in Fig.~\ref{fig:coll_byk_h+} (b) to about 6\%. 
We can thus identify the first peak in Fig.~\ref{fig:coll_byk_h+} (b) as the 
fundamental quadrupole mode of the final equilibrium star. 
Similar to the case of the quasi-radial mode, the difference in the 
frequencies is due mainly to rotation. 

The above results show that the gravitational waveform produced by the 
simplified collapse model is dominated by two oscillation modes. 
By studying the oscillations of a nonrotating star model that has  
the same central density as the final equilibrium star resulting from the 
collapse, we demonstrated that the two oscillation modes 
are the fundamental quasi-radial and quadrupole modes of the final star. 
This conclusion does not depend on the EOS of the collapse model.
We have repeated the above simplified collapse model and mode-identification
analysis with a different polytropic constant $k=48$ (i.e., a 20\% reduction of 
the initial value $k_0=60$) and found that the conclusion still holds. 
We conclude that the excitation of the fundamental quasi-radial and 
quadrupole modes is a general feature of the phase-transition-induced collapse 
of a rapidly rotating neutron star, which does not depend much on the 
EOS model. 
In particular, we show that these two modes are excited in all collapse 
scenarios based on the mixed-phase EOS model.

\subsection{Numerical models}
\label{sec:num_models}

In this section, we present the details of the phase-transition induced 
collapse models based on the mixed phase EOS. 
To generate initial equilibrium neutron star models for this study, we 
solve the time-independent hydrodynamics equations using the self-consistent 
field technique developed by \citet{hac86}.
The initial neutron star is uniformly rotating and modeled by a polytropic 
EOS $P=k_0\rho^{\Gamma_0}$. 
We take $\Gamma_0=2$ and $k_0=60$ in our code units $G=c=M_{\odot}=1$.
Unless otherwise noted, we use $129^3$ Cartesian grid points with 
$\Delta x=\Delta y=\Delta z=0.56$ km. We choose the $z$ axis to be the 
rotation axis in our simulations.

The collapse of the equilibrium neutron star is 
induced by reducing the pressure via the use of the mixed phase EOS
(see~Eq.~(\ref{eq:mixed_EOS})). As discussed in Sec.~\ref{sec:Eos_mixed}, 
we choose the effective adiabatic index $\Gamma$ as the only free parameter 
of the mixed phase EOS. 
We have performed simulations using $\Gamma=1.95$, $1.85$, 
and $1.75$. 
Besides the effect of $\Gamma$, we have also compared the 
gravitational waveforms due to the difference in masses and spin rates of the 
initial neutron star models. 
We have performed the seven specific collapse models listed in Table 1.
In the table we list the initial central densities $\rho_{\rm c}$, rotation 
periods $P$, ratios of the rotational kinetic energy to the potential energy
$T/|W|$, baryonic masses $M$, equatorial radii $R_{\rm e}$, polar 
radii $R_{\rm p}$, and the values of $\Gamma$ for the collapse models. 
Note that we are interested in neutron star models with baryonic 
masses $\gtrsim 2 M_{\odot}$, since it is expected that accreting neutron 
stars with masses over $2 M_{\odot}$ in LMXB are most likely 
to lead to a phase transition \citep{cheng96}.

Model R is a reference model to which the other models are compared. 
The initial neutron star of model R has a (baryonic) mass 2.2 $M_{\odot}$. 
The rotation period is 1.2 ms. The equatorial radius $R_{\rm e}=17.95$ km. 
The ratio of the polar to equatorial radii is 0.69. We take $\Gamma=1.85$
for this collapse model. 
Models G1.95 and G1.75 have the same initial neutron star configuration 
as that of model R. But their $\Gamma$ values are different. Model G1.95 
(G1.75) has a value $\Gamma=1.95\ (1.75)$. On the other hand, 
models M2.0 and M2.4 have different masses compared to model R. 
Model M2.0 (M2.4) has a mass 2.0 (2.4) $M_{\odot}$. 
Finally, models P1.4 and P1.6 have initial rotation period 1.4 and 1.6 ms 
respectively, with the same baryonic mass 2.2 $M_{\odot}$ as the reference 
model R.

\subsection{Collapse and gravitational waveforms}
\label{sec:waveforms}

\begin{figure}
\includegraphics*[width=6.8cm]{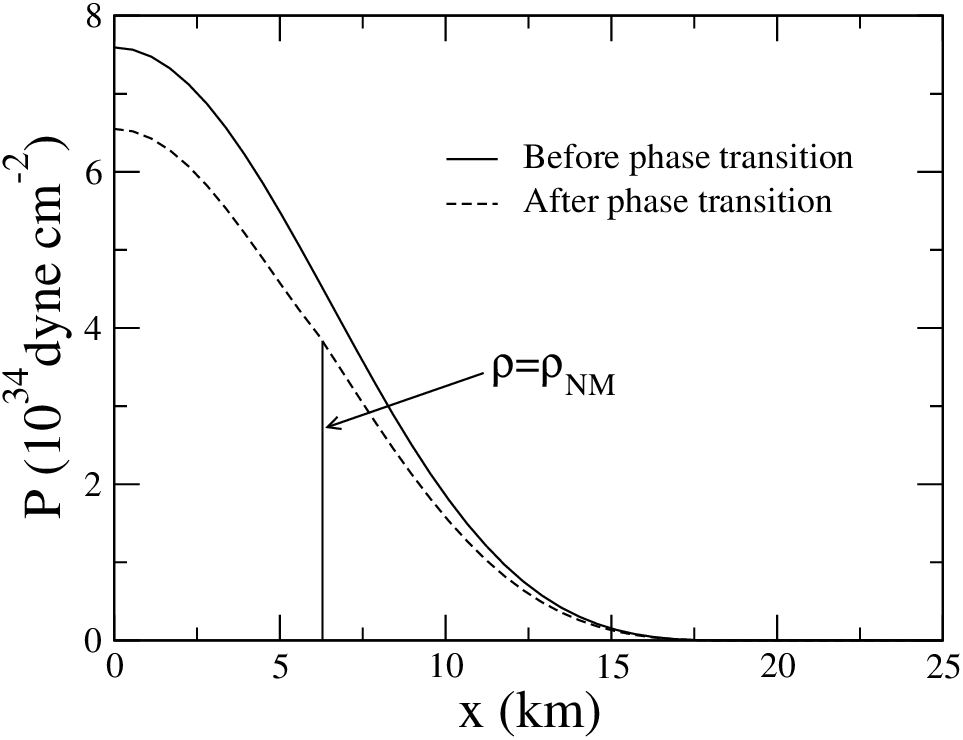}
\caption{Model R. Pressure profiles along the $x$-axis of the initial 
neutron star model (solid line) and that immediately after the phase 
transition (dashed line). The interface between the mixed phase and 
normal nuclear matter is marked by the vertical line $\rho=\rho_{\rm NM}$.}
\label{fig:press_ini_profile}
\end{figure}

Fig.~\ref{fig:press_ini_profile} shows the pressure profiles along the 
$x$-axis for the initial neutron star (solid line) and those immediately
after the phase transition (dashed line) of the reference model R. The 
dashed line in Fig.~\ref{fig:press_ini_profile} corresponds to the 
initial state of the pressure profile for the dynamical evolution. 
Note that the outer boundary of the computational domain is at 
$x= 35.89$ km ($\approx 2R_{\rm e}$) , although we show only up to 
$x=25$ km in the figure. 
The interface between the mixed phase and normal nuclear matter is marked 
by the vertical line $\rho=\rho_{\rm NM}$.
Due to the reduction of the pressure after the phase transition, 
the star undergoes a gravitational collapse. 
The pressure gradient will balance the gravitational 
force, halting the collapse, when the core of the star reaches a high enough density.  
The star subsequently goes into an oscillation phase.

We note that the maximum densities achieved in some of the cases studied reach
$2GM_{\rm c}/R_{\rm c}c^2 \sim 0.5$, where $M_{\rm c}$ is the mass enclosed in the core
region with radius $R_{\rm c}$ defined by the point $\rho = \rho_{\rm NM}$.
This suggests that while it is still consistent to use Newtonian mechanics to study the 
collapse, the phase-transition induced collapse is getting quite close to the point of 
collapsing to a black hole, and an examination of this point using general relativistic 
simulations is called for.  We will revisit this point in the next paper
with the full set of the Einstein equations coupled to the relativistic hydrodynamics equations. 

\begin{figure}
\includegraphics*[width=6.8cm]{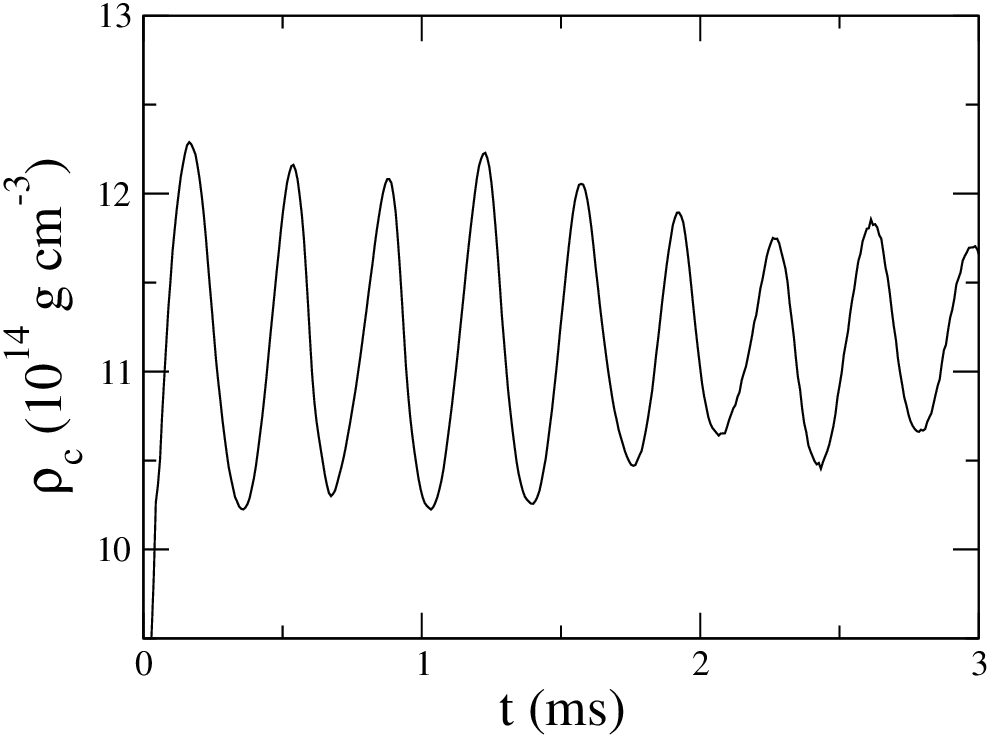}
\caption{Time evolution of the central density for the collapse model R.}
\label{fig:modelR_rhoc}
\end{figure}

In Fig.~\ref{fig:modelR_rhoc} we show the time evolution of the central 
density for the model R. It is seen that the collapse continues until 
the central density rises up to $1.23\times 10^{15}\ {\rm g\ cm^{-3} }$, 
at which point the collapsing star bounces back and oscillations 
are excited subsequently. 

\begin{figure*}
  \begin{minipage}{0.3\linewidth}
        \centering
        \includegraphics*[width=5.5cm]{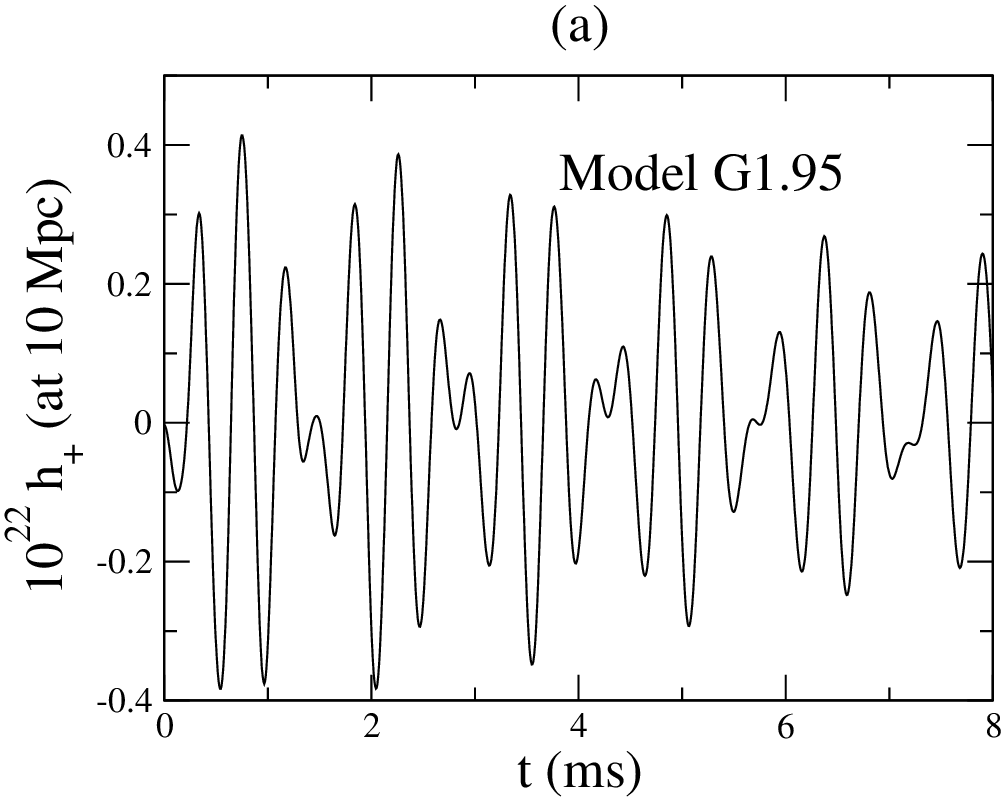}
  \end{minipage}%
  \begin{minipage}{0.4\linewidth}
        \centering
        \includegraphics*[width=5.0cm]{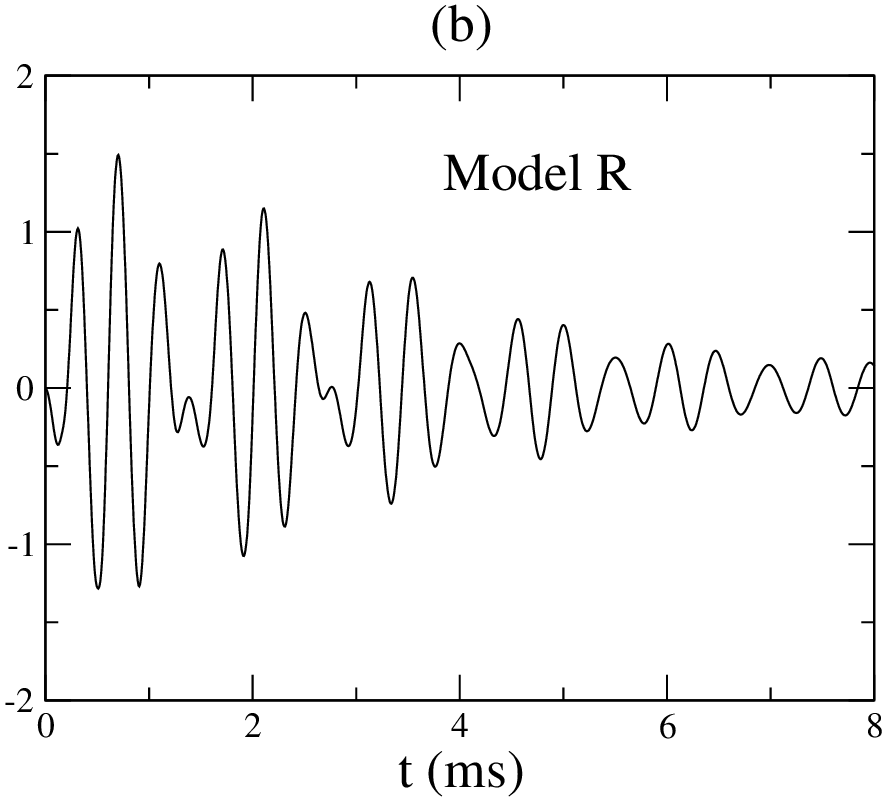}
  \end{minipage}%
  \begin{minipage}{0.3\linewidth}
        \centering
        \includegraphics*[width=5.0cm]{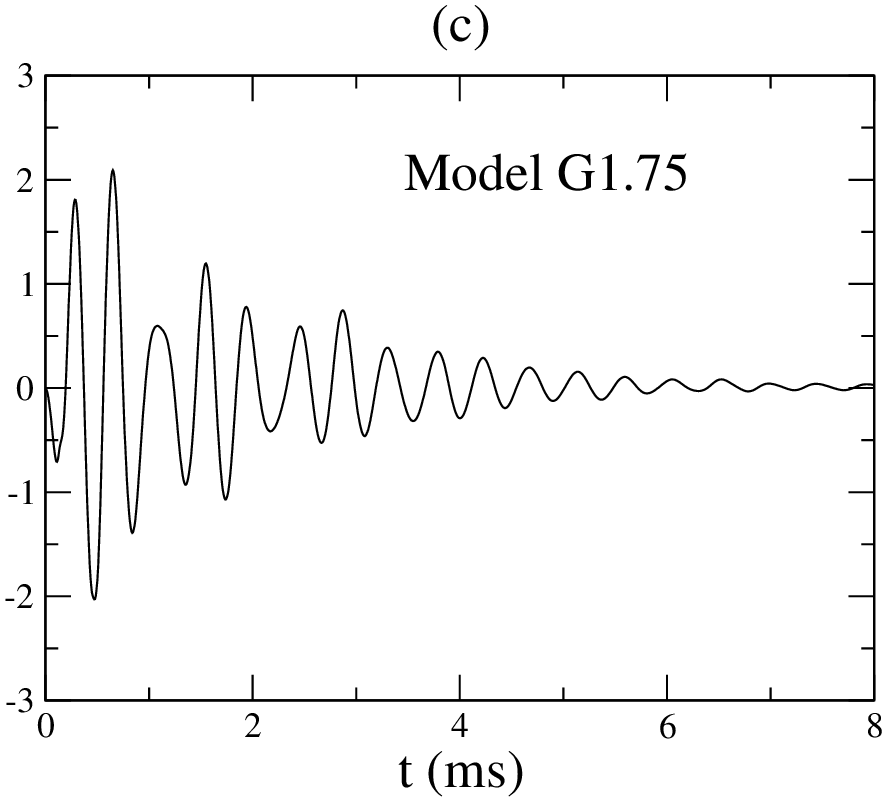}        
  \end{minipage}
   \caption{Gravitational waveforms $h_+$ on the $x$-axis 
            for models G1.95 (a), R (b), and G1.75 (c). 
            The source is assumed to locate at a distance 10 Mpc in all 
            cases.}  
    \label{fig:modelGamma_waveform}
\end{figure*}

The coupling between the star's rotation and oscillations leads to a rapidly 
changing quadrupole mass moment and hence a strong gravitational wave signal. 
In Figs.~\ref{fig:modelGamma_waveform} (a)-(c), we show the time evolution of 
the gravitational waveforms $h_+$ on the $x$ axis for models G1.95, R, and 
G1.75. 
As listed in Table 1, the only difference among the three models in 
Figs.~\ref{fig:modelGamma_waveform} (a)-(c) 
is the effective adiabatic index $\Gamma$ for the nuclear 
matter after phase transition. The figures show that the peak value of the 
gravitational wave amplitude increases as the value of $\Gamma$ decreases.
This is expected, since a smaller value of $\Gamma$ corresponds 
to a larger reduction of the pressure. 
The collapsing star reaches a larger 
compactness before it bounces back, and hence the peak value of the 
gravitational wave amplitude is larger. 
We have also calculated the angle-averaged waveforms 
(see Sec.~\ref{sec:G_wave}) for models G1.95, R, and G1.75. 
We found that the peak value of $< h_+^2 >^{1/2}$ rises from 
$0.3\times 10^{-22}$ to $1.53\times 10^{-22}$ when $\Gamma$ decreases 
from 1.95 (model G1.95) to 1.75 (model G1.75).

Figs.~\ref{fig:modelGamma_waveform} (a)-(c) show that the gravitational wave 
signals for the three models are qualitatively similar, except that for 
models R and G1.75 the wave amplitudes drop significantly within 8 ms. 
The damping of the gravitational wave amplitudes is mostly 
due to the development of differential rotations; and partly due to the formation of 
shock waves leading to the ejection of matter at the stellar surface.

We first consider the collapse model G1.75 and estimate the amount of energy 
available for stellar oscillations triggered by the collapse. 
We note that at any instant the total kinetic energy $T = {1\over 2}\int \rho v^2 d^3x$ 
of the star consists mainly of the rotational energy (uniform and 
differential rotation) $T_{\rm rot}$, with the rest associated with oscillations.
We estimate the maximum amount of energy available for oscillations by the difference 
between $T$ and $T_{\rm rot}$ when the central density reaches the first maximum.

\begin{figure}
\includegraphics*[width=6.5cm]{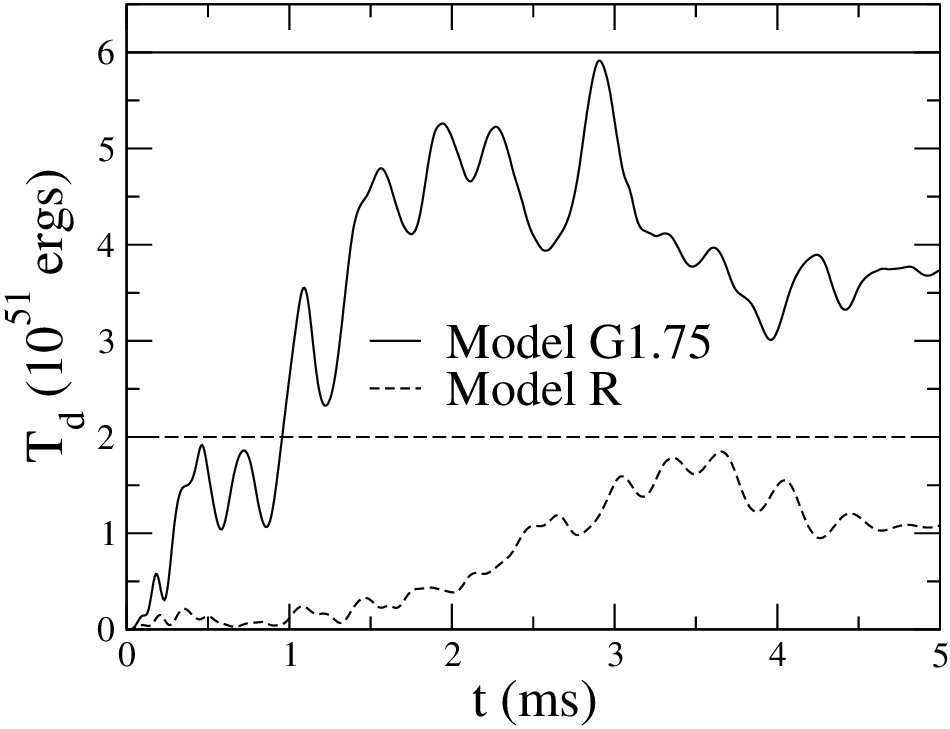}
\caption{Time evolution of the kinetic energy associated with differential rotation 
$T_{\rm d}$. The horizontal lines correspond to the level of the maximum 
oscillation energy (see text) triggered by the collapse for models G1.75 
(solid line) and R (dashed line).}
\label{fig:diff_rot_ke}
\end{figure}

For model G1.75, the central density reaches the first maximum at $t=0.149$ ms, 
and $T=7.3\times 10^{52}$ ergs at that instant. 
The rotational energy is mainly due to uniform rotation at that point 
(see Fig.~\ref{fig:diff_rot_ke}) and is given by 
$T_{\rm rot}= J\bar{\Omega}/2 = 6.7\times 10^{52}$ ergs, where $J$ is the (conserved) total 
angular momentum and $\bar{\Omega}$ is the average angular velocity of the star at that time. 
The oscillation energy as defined is therefore approximately 
$T - T_{\rm rot} \approx 6\times 10^{51}$ ergs. 
Carrying out the same analysis, we find that the oscillation energy for the 
collapse model R is $\approx 2\times 10^{51}$ ergs.

To show that the energy associated with oscillations is mostly converted into 
the energy associated with differential rotation, we define the kinetic energy associated 
with differential rotation by 
\begin{equation}
T_{\rm d} \equiv {1\over 2} \int \rho ( v_{\phi} - \bar{v}_{\phi} )^2 d^3x ,
\end{equation}
where $\bar{v}_{\phi} = \bar{\Omega}(x^2 + y^2)^{1/2}$. For a uniformly 
rotating star, $T_{\rm d}$ vanishes identically. 
(Note that the sum of the energy associated with differential rotation as defined  
and the energy associated with uniform rotation does not equal to the total rotational 
energy.) 
In Fig.~\ref{fig:diff_rot_ke}, we plot $T_{\rm d}$ versus time for models G1.75 and R. 
The horizontal lines correspond to the level of the maximum oscillation energy 
calculated above for models G1.75 (solid) and R (dashed).
Fig.~\ref{fig:diff_rot_ke} shows that, for both models G1.75 and R, $T_{\rm d}$ 
grows rapidly to a level comparable to the oscillation energy by $t\approx 3$ ms.
As can be seen from Figs.~\ref{fig:modelGamma_waveform} (b)-(c), the amplitude 
$h_+$ has dropped significantly by $t\approx 3$ ms.

The above consideration suggests that the strong damping of the oscillations 
seen in models G1.75 and R can be attributed to the growth of differential
rotation during the evolution. Most of the oscillation energy is converted 
to the kinetic energy associated with differential rotation. 
Differential rotation will be damped by viscosity (not modeled in our simulations).

\begin{figure}
\includegraphics*[width=6.8cm]{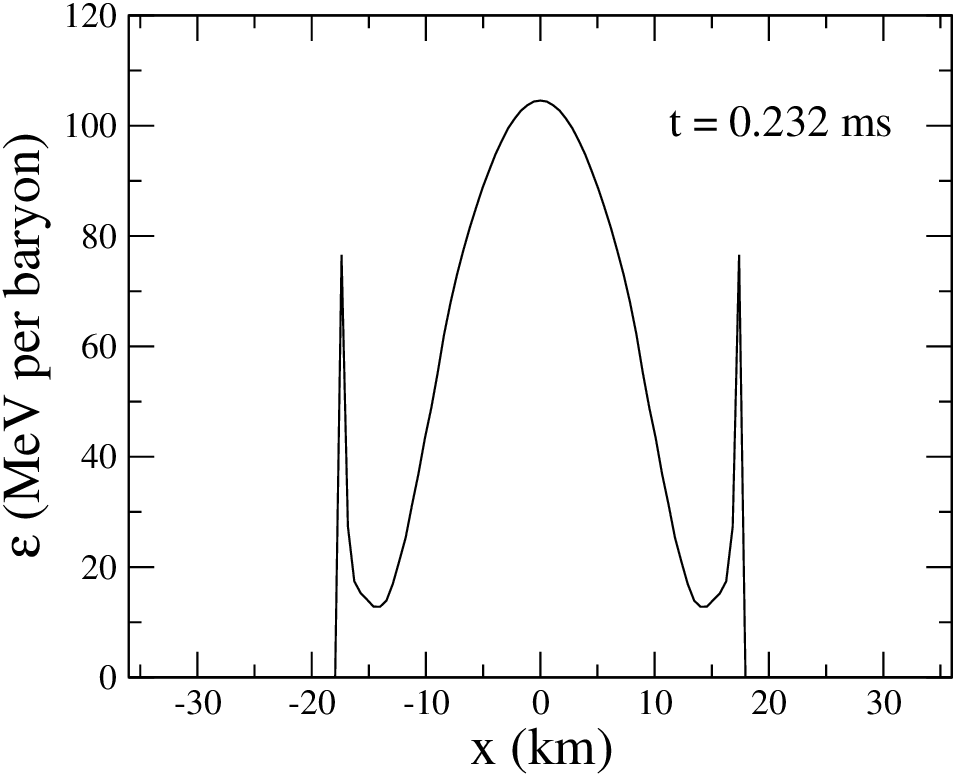}
\caption{Profile of the specific internal energy $\epsilon$ (expressed 
as MeV per baryon) along the $x$-axis for model G1.75 at time $t=0.232$ ms.}
\label{fig:modelG1.75_shock}
\end{figure}

\begin{figure}
\includegraphics*[width=6.8cm]{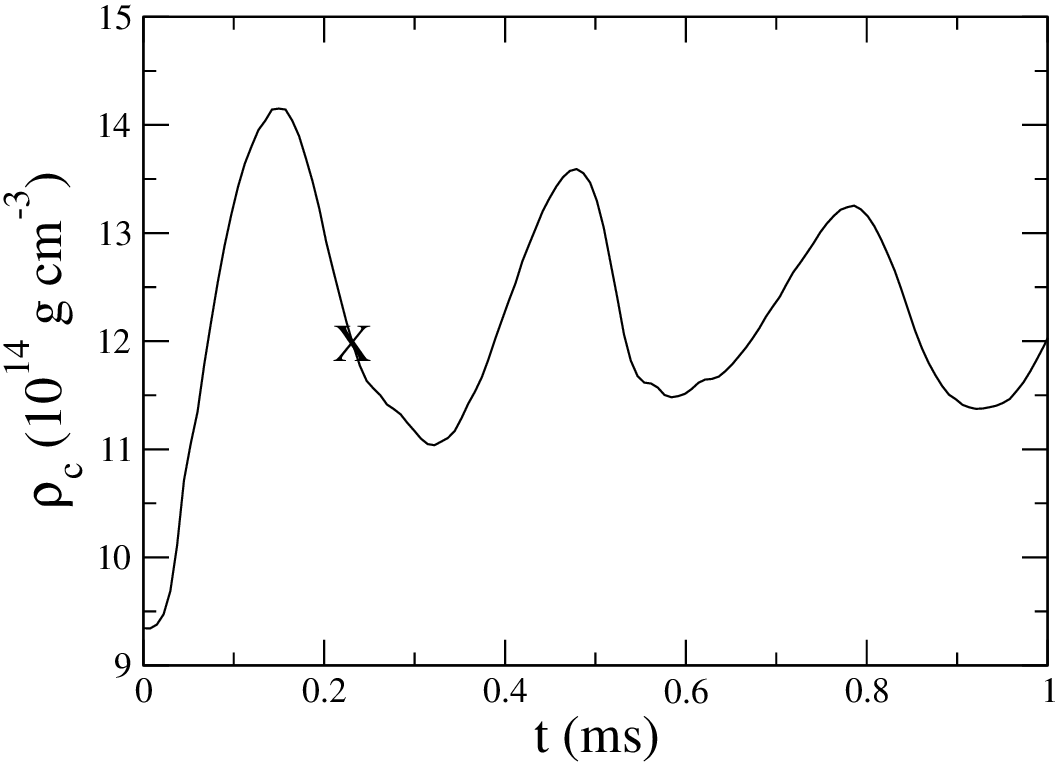}
\caption{Time evolution of the central density of model G1.75. The 
point marked by an ``X'' represents the point when the shock shown in 
Fig.~\ref{fig:modelG1.75_shock} is captured. }
\label{fig:modelG1.75_rhoc}
\end{figure}

\begin{figure}
\includegraphics*[width=6.8cm]{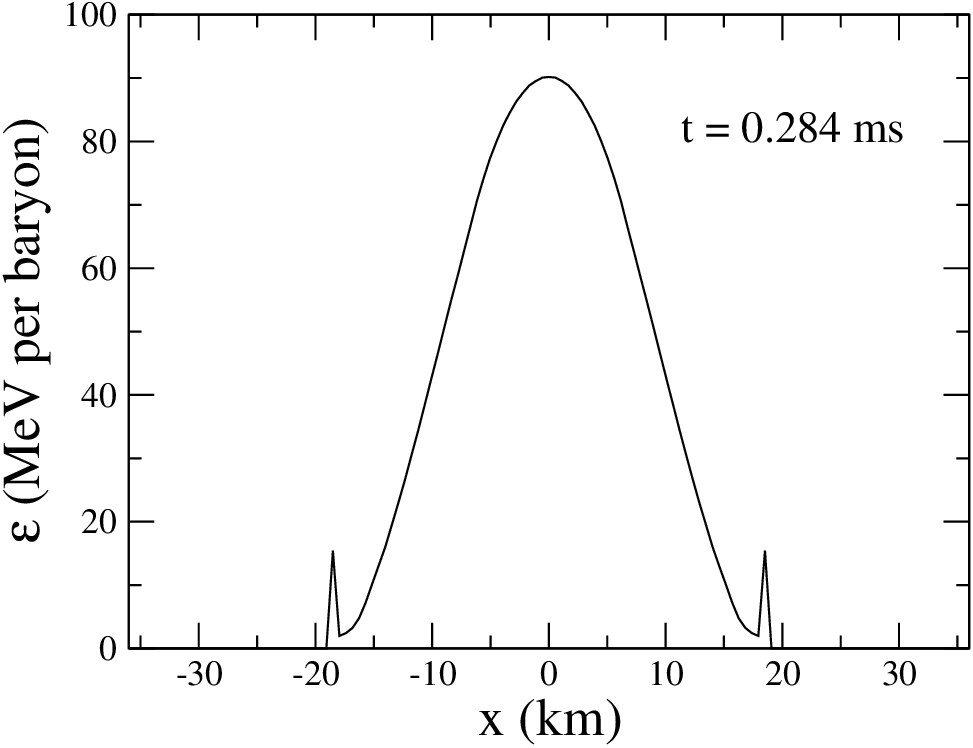}
\caption{Profile of the specific internal energy $\epsilon$ along the 
$x$ axis for model G1.95 at time $t=0.284$ ms.}
\label{fig:modelG1.95_shock}
\end{figure}

Furthermore, some of the oscillation energy is carried away by the shock waves 
formed at the star surface. 
Fig.~\ref{fig:modelG1.75_shock} shows the profile of the specific internal 
energy $\epsilon$ (expressed as mega-electron volts per baryon) along the $x$ axis for 
model G1.75 at time $t=0.232$ ms. 
It is seen that a strong shock wave forms and reaches the surface of the star
shortly before the end of the first oscillation. 
In Fig.~\ref{fig:modelG1.75_rhoc} we plot the time evolution of the central 
density for model G1.75. The point marked by an ``X'' in the figure corresponds to 
the time of Fig.~\ref{fig:modelG1.75_shock}.
Shock waves with weaker amplitudes are formed similarly in each 
oscillation. 
The oscillations are also damped by the shock waves carrying kinetic energy away 
from the star through the ejection of matter from the stellar surface. 
In the model G1.75, we find that the shock waves can dissipate $\approx 10\%$ of 
the oscillation energy by the end of the simulation.
For comparison, Fig.~\ref{fig:modelG1.95_shock} shows the 
first shock wave for model G1.95 at $t=0.284$ ms. The amplitude of the shock 
wave is much smaller.
In this case, the oscillations are less damped by differential rotation and 
shocks, and the gravitational wave signals can last for longer period as 
shown in Fig.~\ref{fig:modelGamma_waveform} (a).

\begin{figure}
\includegraphics*[width=6.8cm]{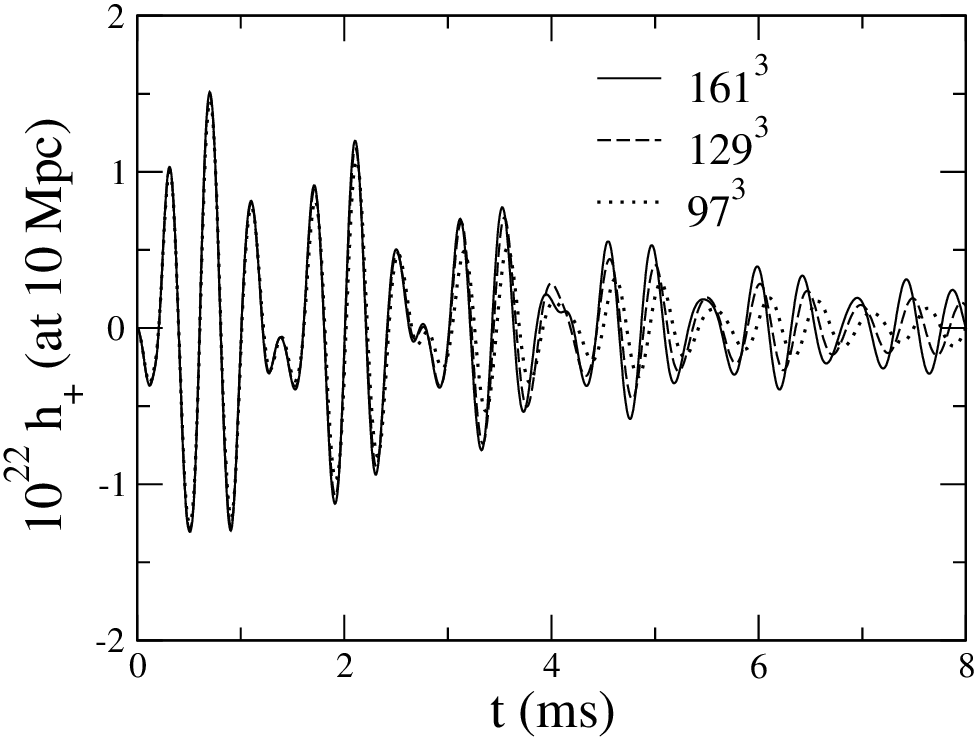}
\caption{Convergence of the gravitational waveform $h_+$, at three 
different resolutions ($97^3$, $129^3$, and $161^3$), for the reference 
model R.}
\label{fig:modelR_h+_convg}
\end{figure}

It should be noted that a small part of the oscillation energy is also lost in 
numerical damping arising from a finite differencing error, which depends on the grid 
resolution used in the simulation. As we see in Fig.~\ref{fig:modelR_h+_convg}, 
for the resolutions used in the simulations, numerical damping is important only for 
the late-time evolution, $t\gtrsim 4$ ms, at which point the oscillations have already 
been damped significantly by the development of differential rotation and shock waves. 
Another indicator of the accuracy of the simulations is that the total energy 
and angular momentum of the system are conserved to about 5\% by the end of the 
simulations for all the collapse models considered in this paper.

\begin{figure}
\includegraphics*[width=6.8cm]{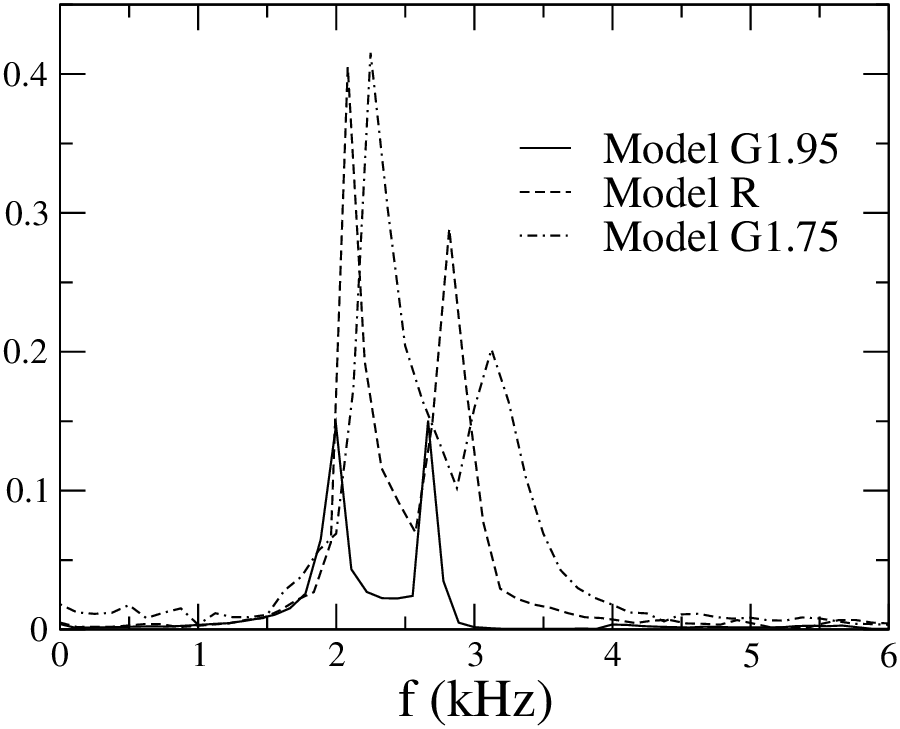}
\caption{Fourier transforms of the wave amplitude $h_+$ for models G1.95, 
R, and G1.75 shown in Fig.~\ref{fig:modelGamma_waveform}.}
\label{fig:modelGamma_h+FT}
\end{figure}

Figs.~\ref{fig:modelGamma_waveform} (a)-(c) show that the gravitational 
waveforms are not dominated by a single oscillation mode: 
the existence of interference between different oscillation modes in the 
gravitational wave signals is clear. 
We plot in Fig.~\ref{fig:modelGamma_h+FT} the Fourier transforms of the time 
evolution of $h_+$ shown in Figs.~\ref{fig:modelGamma_waveform} (a)-(c).
Fig.~\ref{fig:modelGamma_h+FT} shows that the gravitational 
wave signals are dominated by two oscillation modes. 
The frequencies of the oscillation modes increase as the value 
of $\Gamma$ decreases. The difference between the two oscillation frequencies 
$\Delta f$ also increases with a decreasing value of $\Gamma$. It changes from 
$\Delta f = 0.66$ to 0.87 kHz, respectively, 
for models G1.95 and G1.75. 

As we demonstrated in Sec.~\ref{sec:mode_identity}, for the models studied in 
this paper, the higher frequency mode is the fundamental quasi-radial mode of the final 
equilibrium star, while the lower frequency one corresponds to the fundamental quadrupole mode.
In particular, the excitation of these two modes is a general feature of phase-transition-induced 
collapse of a rapidly rotating neutron star that does not depend sensitively on 
the EOS model.

\begin{figure}
\includegraphics*[width=6.8cm]{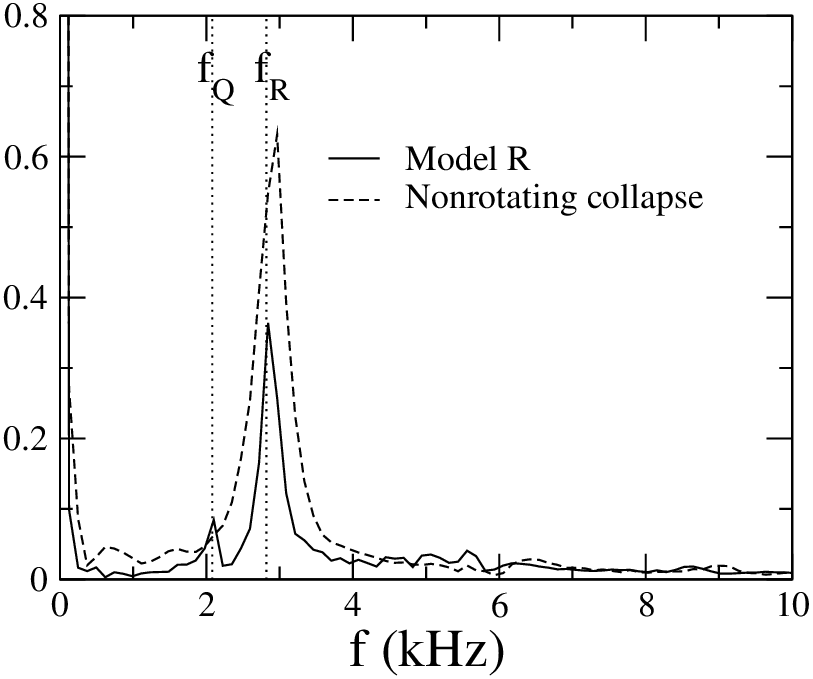}
\caption{Fourier transforms of the time evolutions of the central 
density $\rho_{\rm c}$ for model R and its nonrotating counterpart. 
The positions of the two peaks in the Fourier spectrum of $h_+$ for 
model R are denoted by the vertical dotted lines $f_{\rm Q}$ 
and $f_{\rm R}$. }
\label{fig:modelR_rhoc_compareFT}
\end{figure}

The quasi-radial mode becomes purely radial when the angular velocity of the 
star is zero, in which case it does not radiate gravitational wave. 
In Fig.~\ref{fig:modelR_rhoc_compareFT} we plot the Fourier transforms of 
the time evolution of the central density $\rho_{\rm c}$ for model R 
and its corresponding nonrotating counterpart (defined by having the same 
set of EOS parameters and initial central density as model R).
It is seen that for the nonrotating collapse model (dashed line) the 
spectrum is dominated by the fundamental radial mode at frequency 
2.96 kHz. In the case of rotational collapse (solid line), the 
frequency of the radial mode decreases slightly to $f_{\rm R}=2.82$ kHz. 
The lower peak at $f_{\rm Q}=2.08$ kHz is the fundamental quadrupole mode. 
These two modes are the dominant fluid modes excited in the rotational 
collapse that radiate gravitational waves.

Fig.~\ref{fig:modelGamma_h+FT} also shows that the Fourier amplitudes of 
the higher frequency mode (the quasi-radial mode) for models R and G1.75 are 
smaller than that of the lower frequency one (the quadrupole mode). 
But for model G1.95 the contributions of the two modes to the gravitational 
wave signal are roughly the same. This is due to the severe damping of the 
quasi-radial oscillations by strong shock waves in models R and G1.75. 
In these cases, the quadrupole mode becomes the dominant gravitational-wave-radiating 
mode. On the other hand, the amplitudes of the shock waves formed 
in model G1.95 are much smaller, and the quasi-radial 
oscillations can radiate gravitational waves as strongly as the quadrupole mode.

\begin{figure}
\includegraphics*[width=7.8cm]{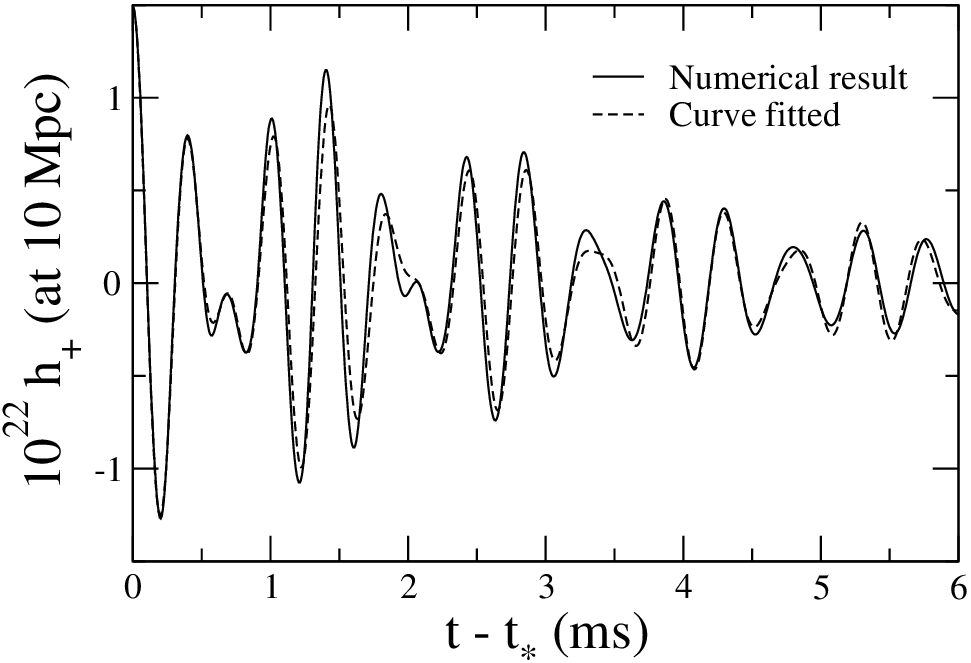}
\caption{Gravitational waveform $h_+$ on the $x$-axis for model R. The solid line is 
the numerical result, while the dashed line is fitted to the function defined in 
Eq. (\ref{eq:h+fit}).}  
\label{fig:h+fit_modelR}
\end{figure}

To further quantity the damping effects, we use the following function to 
fit the gravitational wave amplitude $h_+$:
\begin{equation}
h_+(\bar{t}) = {A_*\over 2}\left[ \exp\left({-{\bar{t}\over \tau_{\rm Q}}}\right)
	\cos (2\pi f_{\rm Q} \bar{t}) + \exp\left({-{\bar{t}\over \tau_{\rm R}}}\right) 
	\cos (2\pi f_{\rm R} \bar{t}) \right] ,
\label{eq:h+fit}
\end{equation}
where $A_*$ is the absolute maximum of the wave amplitude occurring at the time $t=t_*$,
the time coordinate $\bar{t}$ is defined by $\bar{t}=t-t_*$, the $f$ values are the 
frequencies of the modes, the $\tau$'s are the damping time constants of the modes.
For model R, from the numerical waveform and its Fourier transform, 
we have $t_*=0.704$ ms, $A_*=1.49\times 10^{-22}$, $f_{\rm Q}=2.08$ kHz, 
and $f_{\rm R}=2.82$ kHz. 
The two constants $\tau_{\rm Q}$ and $\tau_{\rm R}$ are determined by fitting 
Eq. (\ref{eq:h+fit}) to the numerical data. 
Fig.~\ref{fig:h+fit_modelR} shows the best-fitted curve (dashed line) together with the 
numerical result (solid line). 
It is seen that the fitted curve agrees pretty well with the numerical result. 
The two damping time constants are determined to be
$\tau_{\rm Q}=4.73$ ms and $\tau_{\rm R}=2.56$ ms. 
The smaller value of $\tau_{\rm R}$ compared to $\tau_{\rm Q}$ suggests that the 
quasi-radial mode is damped almost 2 times faster than the quadrupole mode. 
As we have mentioned above, this is due to the strong damping of the quasi-radial 
mode by the shock waves. For comparison, the damping time constants for model G1.75 
are $\tau_{\rm Q}=2.53$ ms and $\tau_{\rm R}=0.42$ ms, while those for model G1.95 
are $\tau_{\rm Q}=10.42$ ms and $\tau_{\rm R}=10.39$ ms. For the latter model, 
the two modes are less damped differential rotation and shock waves, and hence 
they have basically the same (relatively) large damping time constants. 
On the other hand, the modes are damped much faster in model G1.75. In particular,
due to the generation of strong shock waves (see Fig.~\ref{fig:modelG1.75_shock}), 
the quasi-radial mode is damped 6 times faster than the quadrupole mode. 
We have also used Eq. (\ref{eq:h+fit}) to determine the damping time constants for 
the other collapse models studied below. The values are presented in Table 2.

The change in the value of $\Gamma$ at the start of the dynamical evolution 
is to mimic the softening of the nuclear matter 
EOS in the mixed-phase model after the phase transition. 
Our results suggest that the gravitational wave amplitude emitted 
from the collapse and the duration of the wave signal depend quite 
sensitively on the value of $\Gamma$. 
The gravitational wave amplitude can reach a higher peak value if the 
subsequent EOS is significantly softer than that before the phase 
transition. However, the wave will be damped out quickly by the growth 
of differential rotation and the formation of strong shock waves. 
On the other hand, if the stiffness of the  
EOS is comparable to that before the phase transition, the 
gravitational wave signal will last for a significantly longer period. 
A long period could in turn facilitate the detection of the gravitational waves.
We have also performed a simulation with $\Gamma=\Gamma_0=2$, which 
corresponds to the special case where the nuclear matter EOS does not 
change after the phase transition. We find that the emitted 
gravitational wave amplitude is too small to be resolved accurately by 
our grid resolutions. Nevertheless, we estimate that the upper bound of the 
angle-averaged wave amplitude $< h_+^2 >^{1/2}$ is about $10^{-24}$ for a 
source located at 10 Mpc.

In our simulations, we study the effects of the EOS model on the 
gravitational wave signals by varying $\Gamma$, with the bag constant 
fixed to $B^{1/4}=170$ MeV. 
One might also want to vary the value of $B$ in order to study the effects 
of the EOS in details. However, by varying $\Gamma$ alone, we 
have already captured the general feature of what would happen to the 
gravitational wave signals if $B$ was varied instead. 
As suggested by Eq.~(\ref{eq:quark_eos}), a smaller value of $B$ implies 
a stiffer quark-matter EOS, and hence the reduction of pressure after the 
phase transition is smaller. 
This in turn suggests that, if a smaller value of $B$ is used, 
the gravitational waveform of the reference model R 
(see Fig.~\ref{fig:modelGamma_waveform} (b)) will become closer 
to that of model G1.95 (see Fig.~\ref{fig:modelGamma_waveform} (a)): 
the emitted gravitational wave has a smaller amplitude and is less damped. 
On the other hand, if a larger value of $B$ is used, the gravitational wave
signal will be closer to that of model G1.75 
(see Fig.~\ref{fig:modelGamma_waveform} (c)).

Besides the effect of $\Gamma$, we have also studied the gravitational 
wave signals emitted from the collapse process with different initial 
neutron star models. For all cases studied, we find that the gravitational 
waveforms are qualitatively the same as those shown in 
Figs.~\ref{fig:modelGamma_waveform} (a)-(c). In particular, the excitation
of the fundamental quasi-radial and quadrupole modes in the collapse process
does not depend much on the mass and spin rate of the initial 
neutron star model. 

\begin{figure*}
  \begin{minipage}{0.4\linewidth}
        \centering
        \includegraphics*[width=5.5cm]{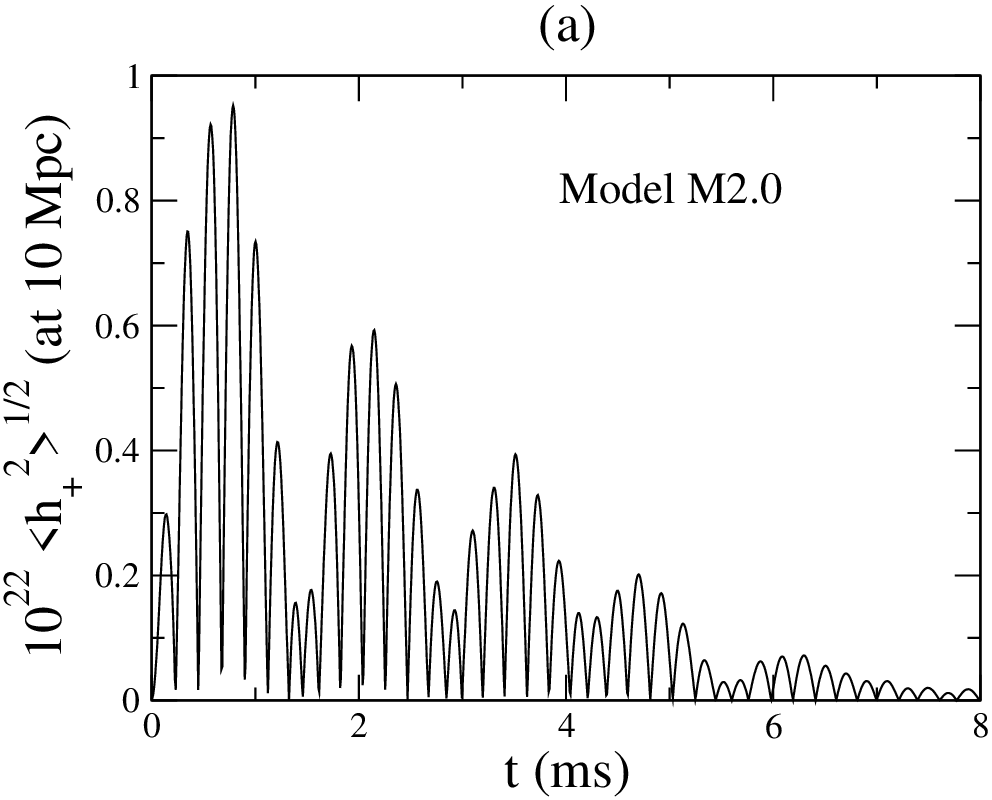}
  \end{minipage}%
  \begin{minipage}{0.4\linewidth}
        \centering
        \includegraphics*[width=5.0cm]{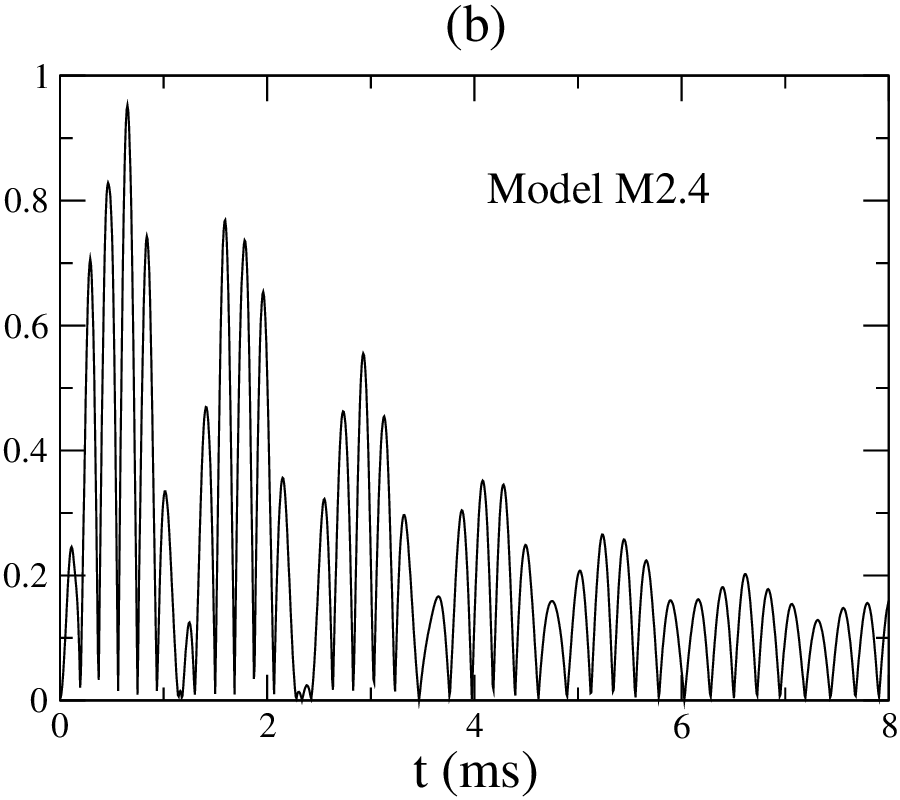}
  \end{minipage}
        \caption{Effect of mass on the gravitational wave signals:
         Angle-averaged wave amplitude $<h_+^2>^{1/2}$ 
         for models M2.0 (a) and M2.4 (b). } 
        \label{fig:modelMass_h+avg}
\end{figure*}

We first study the effect of mass. Figs.~\ref{fig:modelMass_h+avg} (a) and (b) 
show the evolution of the angle-averaged
wave amplitude $< h_+^2 >^{1/2}$, respectively, for models M2.0 and M2.4. 
These models have the same initial rotation period $P=1.2$ ms as the 
reference model R. But note that model M2.0 has the largest $T/|W|$ value 
among all the collapse models (see Table 1). 
We find that the peak value of $< h_+^2 >^{1/2}$ is about 
$0.95\times 10^{-22}$ for a source located at 10 Mpc away in both models. 
The Fourier spectra of the wave amplitude $h_+$ for these models  
also show that the fundamental quasi-radial ($f_{\rm R}$) and 
quadrupole ($f_{\rm Q}$) modes are the dominant contributions to the 
gravitational wave signals. 
The mode frequency  $f_{\rm R}$ ($f_{\rm Q}$) increases from 
2.54 (1.87) kHz to 3.02 (2.21) kHz, when the mass of the star increases
from $2.0$ to $2.4 M_{\odot}$. 
The damping time constants as determined by Eq. (\ref{eq:h+fit}) for model M2.0
are $\tau_{\rm Q}=1.58$ ms and $\tau_{\rm R}=3.23$ ms, while those for model M2.4 
are $\tau_{\rm Q}=4.30$ ms and $\tau_{\rm R}=2.26$ ms. Note that for model M2.0, 
$\tau_{\rm Q}$ is about 2 times smaller than $\tau_{\rm R}$ indicating that 
the quadrupole mode is damped faster.
This is quite different from other collapse models in which $\tau_{\rm Q}$ is 
either comparable to or a few times larger than $\tau_{\rm R}$ (see Table 2).

\begin{figure*}
  \begin{minipage}{0.4\linewidth}
        \centering
        \includegraphics*[width=5.5cm]{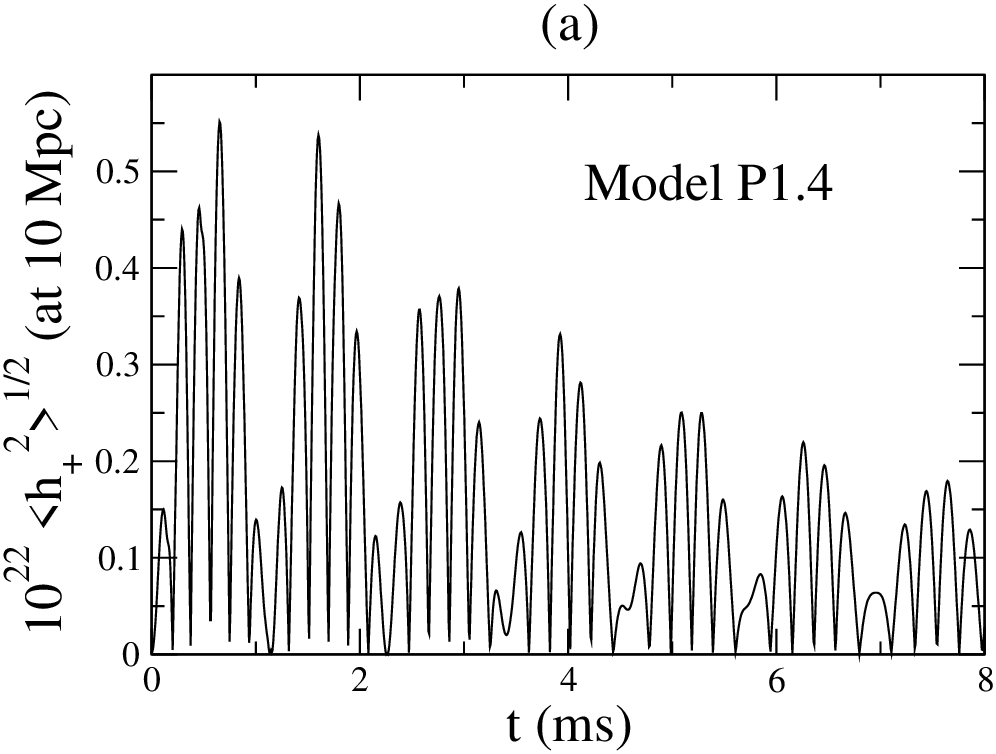}
  \end{minipage}%
  \begin{minipage}{0.4\linewidth}
        \centering
        \includegraphics*[width=5.0cm]{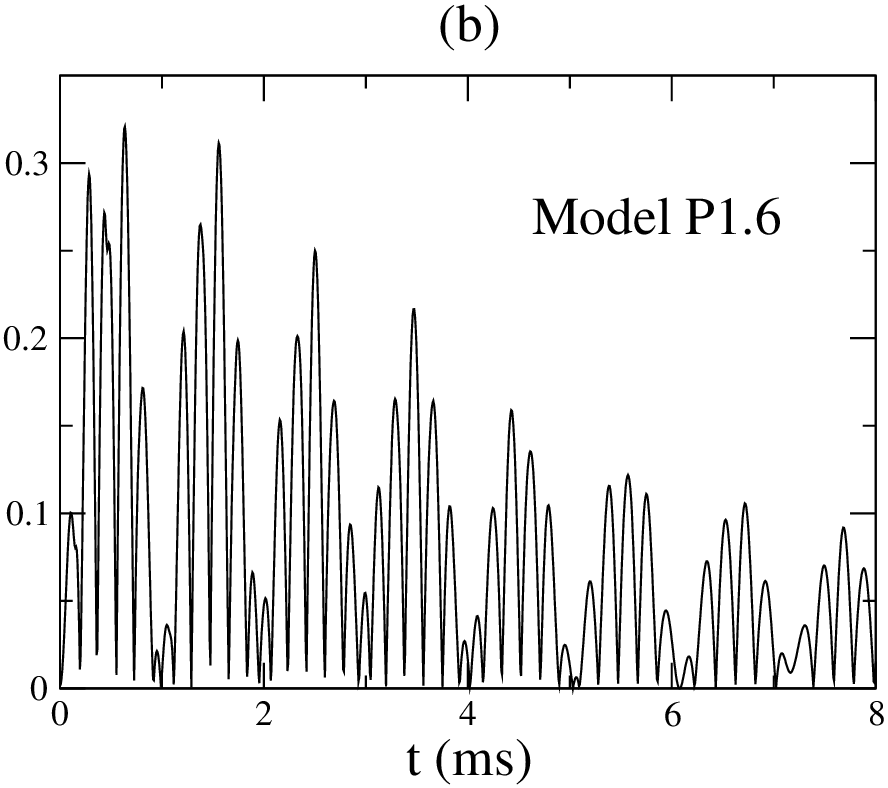}
  \end{minipage}
        \caption{Effect of rotation on the gravitational wave signals.
	Angle-averaged wave amplitude $<h_+^2>^{1/2}$ for models 
	P1.4 (a) and P1.6 (b). These two models have the same baryonic mass,  
         $M=2.2 M_{\odot}$, as the reference model R.}
        \label{fig:modelT_h+avg}
\end{figure*}

\begin{figure}
\includegraphics*[width=6.8cm]{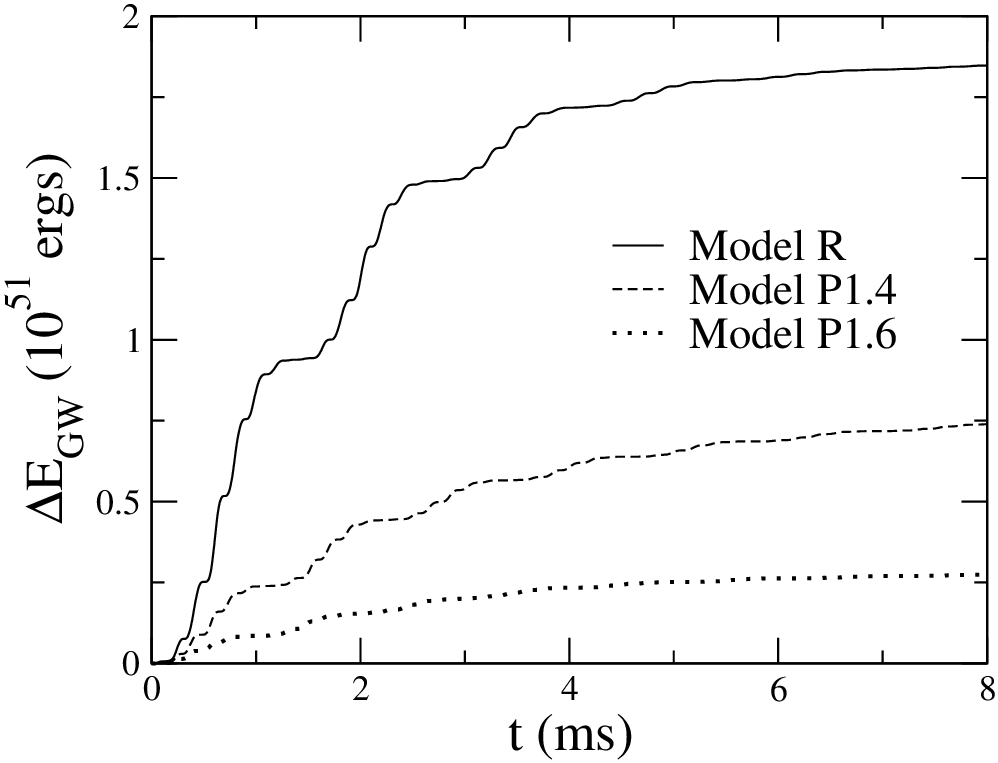}
\caption{Total energy radiated in the form of gravitational waves for 
models R, P1.4, and P1.6.}
\label{fig:modelT_E}
\end{figure}

To study the effect of rotation, we plot in Figs.~\ref{fig:modelT_h+avg} (a) 
and (b) the evolution of $< h_+^2 >^{1/2}$, respectively, for models P1.4 and P1.6. 
These models have the same baryonic mass $M=2.2M_{\odot}$ as the reference model R. 
In contrast to the comparison between models M2.0 and 
M2.4, it is seen that the maximum amplitude of $<h_+^2>^{1/2}$ depends quite  
sensitively on the spin rate of the star. The strong dependence on the 
spin rate can also be seen from the energy radiated $\Delta E_{\rm GW}$
in the form of gravitational waves. Fig.~\ref{fig:modelT_E} shows the total 
energy radiated for models R, P1.4, and P1.6.  
The total energy radiated for model R (with initial rotation period 
$P=1.2$ ms) is $1.85\times 10^{51}$ ergs. It drops down to 
$0.27\times 10^{51}$ ergs for model P1.6. This strong dependence on the 
spin rate can be understood from a dimensional analysis, which suggests that 
the gravitational wave luminosity $dE/dt$ is proportional to $\Omega^6$, 
where $\Omega$ is the angular velocity of the star. 
Form the Fourier transforms of the wave amplitudes $h_+$, we find that 
the frequencies of the two excited modes do not depend sensitively on 
the spin rate of the star (see Table 2).

\begin{figure}
\includegraphics*[width=6.8cm]{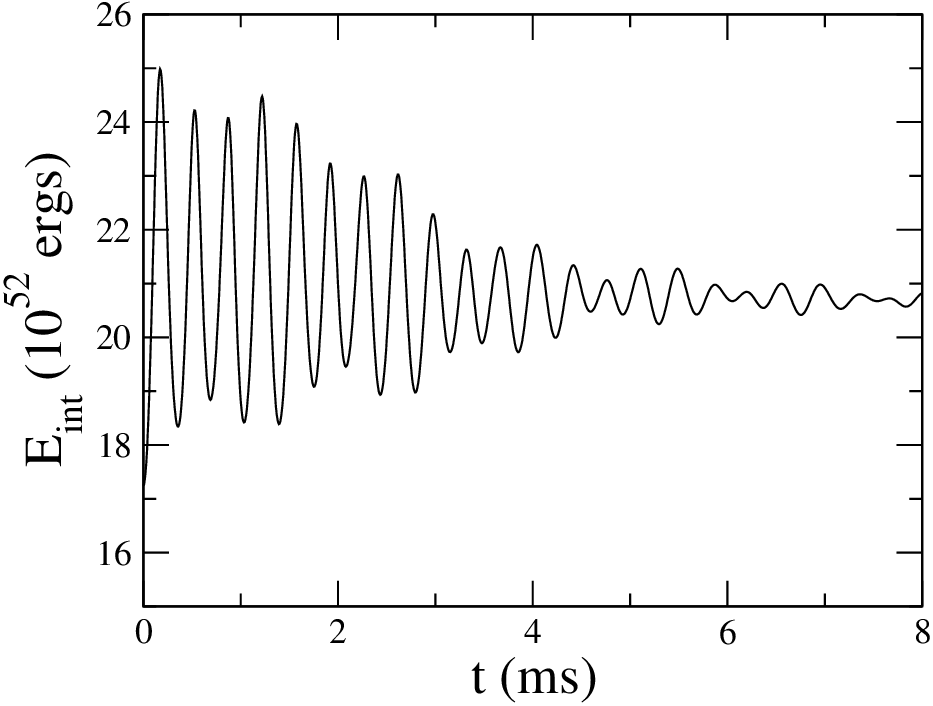}
\caption{Time evolution of the total internal energy of model R.}
\label{fig:modelR_Eint}
\end{figure}

It is also instructive to compare the energy carried away by the 
gravitational waves to the increase in internal energy of the collapsed 
object.  In Fig.~\ref{fig:modelR_Eint} we plot the time evolution of the 
total internal energy $E_{\rm int} = \int \rho\epsilon d^3x$ for the 
collapse model R. After the initial collapse phase, it is seen that 
$E_{\rm int}$ oscillates around its final equilibrium value, 
which we approximate by the time-average value of $E_{\rm int}$ 
in the period 2-6 ms.  The change in the total internal energy in this case is 
$\Delta E_{\rm int} = 3.57\times 10^{52}\ {\rm ergs}$. 
This value corresponds approximately to the difference in the internal energy between 
the final equilibrium configuration and the initial one. On the other hand, 
the energy that would be carried away by the gravitational wave 
$\Delta E_{\rm GW} = 1.85\times 10^{51}\ {\rm ergs}$ amounts to only 
about 5\% of $\Delta E_{\rm int}$. 
(Note that we do not have gravitational radiation reaction in the hydrodynamic 
equations [\ref{eq:rhoeq}]-[\ref{eq:taueq}], and hence the dynamical system is  
conservative.)
 
We have performed the same comparison for all the collapse models 
studied in this paper. We find that the energy ratio 
$\Delta E_{\rm GW}/\Delta E_{\rm int}$ ranges from 0.007 to 0.052. 
In the first few milliseconds covered by our simulations, the remaining 
kinetic energy is mostly in the form of the bulk motion (uniform and 
differential rotation). This kinetic energy will be converted to heat in the 
viscous timescale of the matter in a realistic collapse, longer than that 
covered by our simulations.

In Table 2 we summarize the results of the seven rotational collapse models 
studied in this paper. In the table, $f_{\rm Q}$ ($\tau_{\rm Q}$)
and $f_{\rm R}$ ($\tau_{\rm R}$) are, respectively, the frequencies 
(damping time constants) of the fundamental quadrupole and quasi-radial 
modes obtained from the gravitational wave signals, 
$< h^2_+ >_{\rm peak}^{1/2}$ is the peak value of the angle-averaged wave 
amplitude for a source at 10 Mpc, and  
$\Delta E_{\rm GW}/M c^2$ is the ratio of the total energy emitted in the form of 
gravitational wave to the rest-mass energy of the star. 
To facilitate the comparison between our Newtonian results and future relativistic simulations,  
we also list in the table the angular frequency ($\omega$) of the modes by the dimensionless 
quantity $\omega M$ in units of $G=c=1$, where $M$ is the mass of the star.

To end this section, we demonstrate the accuracy of our numerical results
by showing the convergence of the gravitational waveforms as we increase 
the grid-point resolution. In Fig.~\ref{fig:modelR_h+_convg} we plot 
the evolution of $h_+$ on the $x$ axis for the reference model R with 
three different grid resolutions $97^3$, $129^3$, and $161^3$. 
Fig.~\ref{fig:modelR_h+_convg} shows that the gravitational waveform 
converges very well as we increase the resolution. 
In particular, we see that the three different resolutions 
agree to high accuracy during the initial evolution. 
We conclude that the grid resolution $129^3$ used for all collapse 
models in this paper is in the convergence regime for the physics 
results reported in this paper.

\section{Summary and Discussion}
\label{sec:conclusion}

We have performed dynamical evolutions of phase-transition-induced collapse of 
rapidly rotating neutron stars using a three-dimensional Newtonian 
hydrodynamic code.  
We assume that once a seed of strange quark matter is formed inside the core 
of a neutron star, the star will convert to a quark star in a 
timescale much shorter than the dynamical timescale of the system. 
The resulting quark star is described by a mixed phase EOS model. 
Due to the reduction of the pressure, the star collapses promptly and 
stellar oscillations are excited. 
We have calculated the emitted gravitational wave signals by the quadrupole 
formula.

In our simulations, we use an ideal-fluid EOS 
$P=(\Gamma-1)\rho\epsilon$ to model the nuclear-matter phase. The MIT bag 
model with a bag constant $B^{1/4}=170$ MeV is used to describe the 
quark-matter phase. We consider 
$\Gamma$ as the only free parameter in our mixed-phase EOS model. 
We have found that the gravitational wave amplitudes depend quite 
sensitively on the value of $\Gamma$. 
The gravitational waves have higher amplitudes if the subsequent EOS of the 
matter is significantly softer than that before the phase transition. 
For our models, the peak value of the angle-averaged waveform 
$< h_+^2 >_{\rm peak}^{1/2}$ can reach the order $10^{-22}$ for a source  
distance of 10 Mpc. 
However, the waves are damped very quickly by the growth of differential 
rotation and the formation of strong shock waves at the stellar surface. 
On the other hand, if the stiffness of the EOS does not change 
much after the phase transition (i.e., $\Gamma\lesssim \Gamma_0$), the fluid 
oscillations are less damped and the emitted gravitational wave signals will 
last for a longer period, albeit with smaller amplitudes. 
Furthermore, for all cases studied in this paper, we find that the energy 
carried away by the gravitational waves amounts to only a few percent or 
less of the increase in internal energy of the collapsed stars. 
This suggests that in a realistic collapse scenario, most of the gravitational 
potential energy released in the collapse will be converted into heat due to viscous 
damping and be carried away by radiation (neutrino and electromagnetic)
in a much longer timescale.

Besides the effect of the EOS model, we have also studied the effects of 
mass and spin rate of the initial neutron star model on the gravitational 
wave signals. We find that for a fixed initial spin rate, the amplitude 
of the gravitational waves does not depend sensitively on the mass 
of the star. On the other hand, rotation of the star 
can enhance the amplitude of the gravitational waves significantly. 
The peak value of the angle-averaged waveform $<h_+^2>^{1/2}$ for a source 
at 10 Mpc can rise from $0.32\times 10^{-22}$ to $1.09\times 10^{-22}$ 
when the initial rotation period decreases from 1.6 ms (for model P1.6) to 
1.2 ms (for model R).

We have found that the gravitational wave emission is dominated by 
two particular oscillation modes excited by the collapse.
By studying a simplified collapse model and the oscillation modes of the 
corresponding final equilibrium star, it is found that the two excited 
fluid modes are the fundamental quasi-radial and quadrupole modes. 
The excitation of these two particular modes is quite general and does not 
depend sensitively on the EOS model. 
Hence, we expect that regardless of the EOS model of the quark star, which is 
not well understood at the moment, the gravitational wave signals emitted 
from the phase-transition induced collapse of rapidly rotating neutron stars 
are dominated by the fundamental quasi-radial and quadrupole modes of the 
final equilibrium stars. 
While it is generally expected that nonradial oscillation modes  
and various instabilities associated with them are the most promising 
sources of gravitational waves emitted from isolated neutron stars \citep{and03}, 
we have shown explicitly that the fundamental quasi-radial modes in rapidly 
rotating compact stars can also lead to significant emission of gravitational 
waves. For some of the collapse models studied in this paper, they can 
contribute to the gravitational wave signals as strongly as the fundamental 
quadrupole modes. 

Our simulations are performed in full three-dimensions without any symmetry assumptions, and 
we do not see any excitation of non axisymmetric modes for the collapse models 
studied in this paper. 
The collapse scenarios that can lead to the development of nonaxisymmetric 
dynamics will be studied in a future paper.

Are the gravitational wave signals from phase-transition induced collapses 
detectable? 
Table 2 shows that the gravitational wave signals emitted from our 
collapse models have frequencies $\sim 2-3$ kHz and peak amplitudes 
$\sim 10^{-22}$ for a source distance of 10 Mpc. 
The relatively high frequency range puts the signals below the sensitivity 
of LIGO. 
However, for sources located within our Galaxy ($\sim 10$ kpc), the signals 
attain values well above the noise level of LIGO. 
With our conclusion that the waveforms are sensitive to the EOS, 
the detection of such gravitational wave signals would provide  
information regarding the EOS of newly born quark stars.

Finally, we conclude with several remarks. (1) Although our numerical 
simulations are performed in Newtonian gravity, we expect the same two 
dominant fluid modes to be excited in phase-transition induced collapse 
in full general relativity (GR). 
(2) For the models studied in this paper, we checked that at the maximum contraction point 
although the collapse (in Newtonian dynamics) is not strong enough to produce a black hole, 
gravity is so strong that general relativistic effects could be significant, and we will 
revisit this point with a full GR simulation in the next paper.
(3) While the MIT bag model is only a simple phenomenological model for quark matter, 
we believe the main features found in this paper is generic to phase-transition 
induced collapse. 
(4) We have ignored the neutrino emission in our study.
This is consistent with the maximum internal energy achieved in the collapse.
The dynamics in the first few milliseconds is not strongly affected by 
neutrino emission, which, however, will be essential in the subsequent cooling (in the 
timescale of seconds or more; \citet{sha83}) of the collapsed object.
(5) We have also ignored the effect of internal fluid dissipation in our 
simulations. The shear viscosity damping timescale of strange quark matter is 
comparable to that of normal nuclear matter and is on the order of $10^8$ s 
for a star model with mass $M=1.4M_{\odot}$, radius $R=10$ km, and temperature 
$T=10^9$ K. The bulk viscosity damping timescale for the same star model is on the 
order of seconds (assuming a rotation period of 1 ms and an $s$-quark 
mass $m_{s}=100$ MeV; \citet{mad00}). 
We thus conclude that viscosity damping is not important to the dynamics in 
the timescale $\sim 10$ ms studied in this paper.


\begin{acknowledgments}
We would like to thank Nils Andersson, Ian Hawke, and Nikolaos Stergioulas for useful 
discussions, and J\'{e}r\^{o}me Novak for comments on the manuscript. 
The Newtonian hydrodynamics code used in this work is written and supported by Philip 
Gressman and Lap-Ming Lin. This work is supported in parts by a RGC grant of 
the Hong Kong Government, NSF Grant Phy 99-79985 (KDI Astrophysics Simulation
Collaboratory
Project), and NSF NRAC MCA93S025. L. M. L. acknowledges support from a Croucher
Foundation fellowship. All computations were performed on the NCSA TeraGrid Linux
cluster. 

\end{acknowledgments}





\clearpage

\clearpage

\clearpage

\begin{deluxetable}{c c c c c c c c}
 \tablecaption{Phase-transition-induced collapse models}

 \tablehead{ 
   \colhead{Model\tablenotemark{a} }
     & \colhead{$\rho_{\rm c}\ {\rm (10^{14} g\ cm^{-3})}$} 
     & \colhead{$P\ {\rm (ms)}$} 
     & \colhead{$T/|W|$}	
     & \colhead{$M(M_{\odot})$}  
     & \colhead{$R_{\rm e}\ {\rm (km)}$}  
     & \colhead{$R_{\rm p}\ {\rm (km)}$} 
     & \colhead{$\Gamma$} }
 \startdata
   R  & 9.34 & 1.2 & 0.08 & 2.2 & 17.95 & 12.34 & 1.85 \\
   G1.95  & 9.34 & 1.2 & 0.08 & 2.2 & 17.95 & 12.34 & 1.95 \\
   G1.75 & 9.34 & 1.2 & 0.08 & 2.2 & 17.95 & 12.34 & 1.75 \\
   M2.0 & 7.73 & 1.2 & 0.11  & 2.0 & 21.31 & 11.78 & 1.85 \\
   M2.4 & 10.55 & 1.2 & 0.07 & 2.4 & 17.39 & 12.34 & 1.85 \\
   P1.4 & 10.30 & 1.4 & 0.05 & 2.2 & 16.26 & 12.90 & 1.85 \\
   P1.6 & 10.69 & 1.6 & 0.03 & 2.2 & 15.14 & 13.46 & 1.85 \\
   \enddata
\tablenotetext{a}{In all cases, the initial neutron stars are modeled by 
a polytropic EOS $P=k_0\rho^{\Gamma_0}$, with $k_0=60$ and $\Gamma_0=2$.
The other parameters for the mixed phase EOS are 
$B^{1/4}=170$ MeV, $\rho_{\rm NM}=7.25\times 10^{14}\ {\rm g\ cm^{-3}}$, 
and $\rho_{\rm QM}=2.52\times 10^{15}\ {\rm g\ cm^{-3}}$. 
Model R is the reference model with which the other collapse models are 
compared.}
\end{deluxetable}

\begin{deluxetable}{c c c c c c c}
 \tablecaption{Gravitational wave signals from the phase-transition 
               induced collapse models}

 \tablehead{
    \colhead{Model}
  & \colhead{$f_{\rm Q}$ ($\omega_{\rm Q}M$)\tablenotemark{a}}
  & \colhead{$\tau_{\rm Q}$\tablenotemark{b}} 
  & \colhead{$f_{\rm R}$ ($\omega_{\rm R}M$)} 
  & \colhead{$\tau_{\rm R}$}	
  & \colhead{$<h_+^2>_{\rm peak}^{1/2}\ (10^{-22})$\tablenotemark{c}} 
  & \colhead{$\Delta E_{\rm GW}/M c^2$}  }
 \startdata 
    R  & 2.08 (0.142) & 4.73  & 2.82 (0.192) & 2.56  & 1.09 & $4.7\times 10^{-4}$ \\
 G1.95 & 2.00 (0.136) & 10.42 & 2.66 (0.181) & 10.39 & 0.30 & $7.1\times 10^{-5}$ \\
 G1.75 & 2.25 (0.153) & 2.53  & 3.12 (0.212) & 0.42  & 1.53 & $7.0\times 10^{-4}$ \\
  M2.0 & 1.87 (0.116) & 1.58  & 2.54 (0.157) & 3.23  & 0.95 & $3.1\times 10^{-4}$ \\
  M2.4 & 2.21 (0.164) & 4.30  & 3.02 (0.224) & 2.26  & 0.95 & $3.9\times 10^{-4}$ \\
  T1.4 & 2.18 (0.148) & 3.88  & 3.01 (0.205) & 3.31  & 0.55 & $1.9\times 10^{-4}$ \\
  T1.6 & 2.12 (0.144) & 4.08  & 3.11 (0.212) & 4.56  & 0.32 & $6.9\times 10^{-5}$ \\
  \enddata
\tablenotetext{a}{The frequency $f$ is in kHz. The angular frequency 
$\omega=2\pi f$ is expressed by the dimensionless quantity 
$\omega M$ in units of $G=c=1$, where $M$ is the mass of the star.}
\tablenotetext{b}{The damping time constant $\tau$ is in ms.}
\tablenotetext{c}{The source is assumed to be at a distance 10 Mpc.}
\end{deluxetable}

\end{document}